\documentclass[useAMS,usenatbib]{mn2e}
\usepackage{graphicx}
\usepackage{latexsym}
\usepackage{psfig}
\usepackage{subfigure}
\usepackage{url}
\usepackage{amsmath}

\title{The fragmentation of expanding shells III: Oligarchic accretion and the mass spectrum of fragments}

\author[James E. Dale, Richard W\"unsch, Rowan J. Smith, Anthony Whitworth, Jan Palou\v{s}]{James E. Dale$^{1}$\thanks{E-mail: jim@ig.cas.cz (JED)},Richard W\"unsch$^{1,2}$,Rowan J. Smith$^{3}$,Anthony Whitworth$^{2}$,Jan Palou\v{s}$^{1}$\\
$^{1}$Astronomical Institute, Academy of Sciences of the Czech Republic, Bocni II 1401/2a, 141 31 Praha 4\\
$^{2}$School of Physics and Astronomy, Cardiff University, Queens Buildings, The Parade, Cardiff, CF24 3AA\\
$^{3}$Institut f\"ur Theoretishce Astrophysik, Universit\"at Heidelberg, Albert--\"Uberle--Str. 2, Heidelberg, Germany}

\begin{document}

\pagerange{\pageref{firstpage}--\pageref{lastpage}} \pubyear{2006}

\maketitle

\label{firstpage}

\def\mnras{MNRAS}
\def\apj{ApJ}
\def\aap{A\&A}
\def\apjl{ApJL}
\def\apjs{ApJS}
\def\bain{BAIN}
\def\araa{ARA\&A}
\def\pasp{PASP}
\def\aj{AJ}
\def\pasj{PASJ}
\def\ga{\sim}

\newcommand{\half}{_{_{1\!/\!2}}}
\newcommand{\OO}{_{_{\!0\!}}}

\begin{abstract}
We use SPH simulations to investigate the gravitational fragmentation of expanding
shells through the linear and non--linear regimes. The results are analysed using
spherical harmonic decomposition to capture the initiation of structure during the
linear regime; the potential-based method of Smith et al. (2009) to follow the
development of clumps in the mildly non-linear regime; and sink particles to capture
the properties of the final bound objects during the highly non-linear regime. In
the early, mildly non--linear phase of fragmentation, we find that the clump mass
function still agrees quite well with the mass function predicted by the analytic
model. However, the sink mass function is quite different, in the sense
of being skewed towards high-mass objects. This is because, once the growth of a condensation becomes non-linear, 
it tends to be growing non-competitively from its own essentially separate reservoir; we call this
Oligarchic Accretion.
\end{abstract}


\section{Introduction}
Bubbles and the shells they sweep up are ubiquitous features of the interstellar
medium (ISM), on a wide range of mass and size scales. \cite{2006ApJ...649..759C} and \cite{2007ApJ...670..428C} observed over 600 infrared shells with Spitzer on size scales of $0.1$ to
$1.0\,{\rm pc}$. \cite{2007A&A...463.1227K} catalogued 462 larger--scale loops and
arcs observed in the far infrared and \cite{2005A&A...437..101E} found
$\sim1000$ shell--like structures in the Leiden-Dwingeloo HI survey on scales of
tens to hundreds of parsecs. These shells are of intrinsic interest, since they form
an important part of the structure of the ISM, but they are also interesting from
the point of view of star formation, since it has long been suspected that dense
shells are sites of triggered star formation; hence they form a crucial component in
the self--propagating star formation model \citep{1977ApJ...214..725E}.\\
\indent There exists a considerable body of theoretical work on the fragmentation of
expanding shells, usually in the context of the thin-shell model \citep[e.g][]{1994ApJ...427..384E, 1994A&A...290..421W, 1994MNRAS.268..291W, 2001A&A...374..746W}. The thin-shell model
treats the expanding shell as infinitesimally thin and considers the linear behaviour
of perturbations in the shell surface-density, treating the fragmentation as a
locally two-dimensional process. However, there is no clear consensus on the mass
function of stars produced by this process. \cite{1994MNRAS.268..291W} assume that the
most unstable mode in the shell can be used to characterize the mass function,
concluding that shell fragmentation should preferentially produce high-mass stars.
If this is correct, it strongly supports the self-propagating star formation model,
since each shell produces more than enough O-type stars to trigger the next stellar
generation. However, \cite{2001A&A...374..746W} perform a more detailed
analysis under the assumption that all unstable wavelengths are represented in the
mass function. They derive a fragment mass function from the dispersion relation for
the growth rate of perturbations of different angular wavenumbers, and conclude that
shell fragmentation should produce a power--law mass function not very different
from the canonical Salpeter function.\\
\indent Unambiguous observational evidence for shell fragmentation is difficult to obtain. \cite{2007ApJ...670..428C} find that $\sim18\%$ of their sample of 269 bubbles in the inner Milky Way show evidence of triggered star formation, but attribute this to the shells over--running pre--existing density anomalies, rather than to fragmentation of the shells themsleves.
Observations of shells driven by HII regions by, for example, \cite{2006A&A...446..171Z}, \cite{2006A&A...458..191D} and \cite{2008A&A...482..585D} often detect populations of young stars in the peripheries of bubbles and the stars are often massive. Herschel observations of  Sh2--104 \citep{2010arXiv1005.3070R} and RCW120 \citep{2010arXiv1005.1615Z} seem to imply that shell
fragmentation produces small numbers of large fragments. However, observations by \cite{2008PASJ...60.1297D} of the mass function of CO clumps in the Carina Flare supershell reveal a power--law mass spectrum. It is thus not clear from either an observational or a theoretical point of view how self--gravitating shells should fragment.\\
\indent The present paper is one of a series aimed at resolving these issues by performing
detailed numerical simulations of fragmenting shells. In the first paper (\cite{2009MNRAS.398.1537D}; hereafter Paper I) we studied the fragmentation of expanding shells in the
linear regime using a Smoothed Particle Hydrodynamics (SPH) code and an Adaptive
Mesh Refinement (AMR) code. We studied the growth of perturbations in the shell
surface density and compared the results of the two codes with each other and with
the thin--shell model. We found that the agreement between the two radically
different numerical schemes was excellent. However, we also found that the boundary
conditions applied to the shell were of crucial importance in determining the
mass-spectrum of perturbations. A shell expanding in a vacuum gets thicker as it
expands, and this thickening suppresses fragmentation at short wavelengths, relative
to the predictions of the thin shell model. Conversely, if the thickening of the shell is prevented by
a confining pressure, the shorter wavelengths become unstable too and, for a certain finite value of the pressure,
fragmentation proceeds in good agreement with the thin shell
model.\\
\indent The second paper in this series, \cite{2010arXiv1005.4399W} and hereafter Paper II, presents an alternative to the thin--shell model which allows the boundary conditions on the shell inner and outer surfaces to be included and thus the effects of external pressure to be treated analytically. We find that pressures higher than that required to keep the shell thickness constant extend the range of unstable wavelengths towards lower values. This should in principle result in a fragment mass function with a steeper slope and we investigate this possibility below.\\
\indent In this paper, we extend some of the simulations of Paper I into the non-linear
regime and allow the fragmentation of the shells to proceed to the formation of
bound compact objects. These can be thought of either as protostars or
protoclusters, depending on their masses, allowing us (i) to derive a mass function
for the stellar/cluster populations formed by fragmenting shells; and (ii) to
evaluate the ability of the thin--shell model to predict these mass functions. In
Section 2, we discuss the thin--shell model and the means by which mass functions
may be inferred from it. In Section 3, we explain our numerical methods and in
Section 4, we briefly outline the simulations conducted. Our results are presented
in Section 5 and discussed in Section 6. Finally, we summarize our conclusions in
Section 7.
\section{Mass functions from the thin shell model}
\begin{figure}
\includegraphics[width=0.45\textwidth]{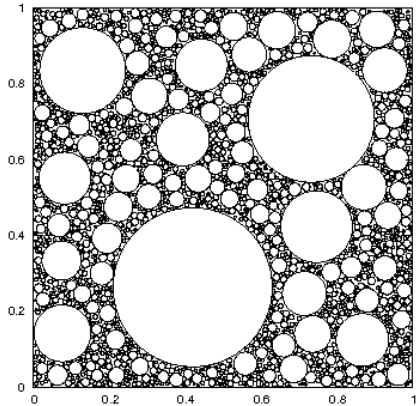}
\caption{Result of an experiment of covering a square with circles uniformly drawn in mass from $\left[ 1.26\times10^{-5},7.9\times10^{-1}\right]$ up to a covering factor of 0.8.}
\label{fig:circles}
\end{figure}

\begin{figure}
\includegraphics[width=0.45\textwidth]{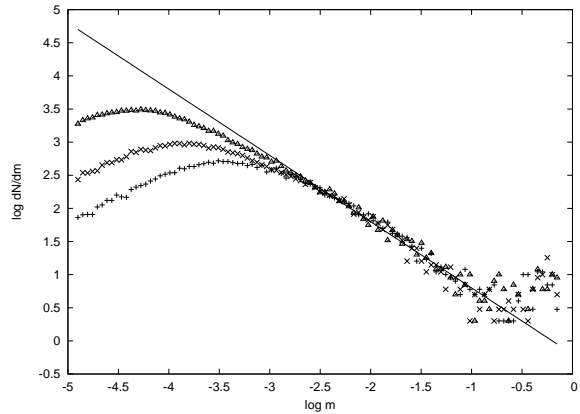}
\caption{Mass functions of circles, with covering factors $f$ of 0.70 (plus signs), 0.75 (crosses) and 0.80 (triangles). The straight line has a logarithmic slope of -2.0.} 
\label{fig:circles_mf}
\end{figure}

For a self--gravitating shell with instantaneous radius, expansion velocity and mean
surface density given by $R$, $V$ and $\Sigma_{0}$, the infinitesimally thin shell model yields an
expression for the growth rate $\omega$ of perturbations in the surface density of
spherical harmonic wavenumber $l$,
\begin{eqnarray} 
\omega(l)=-\frac{AV}{R}+\sqrt{\frac{BV^{2}}{R^{2}}-\frac{c_{s}^{2}l^{2}}{R^{2}}+\frac{2\pi
G\Sigma_{0} l}{R}},
\label{eqn:disp_rel}
\end{eqnarray}
where $c_{s}$ is the sound speed inside the shell, and $A$ and $B$ are constants
whose values depend on whether the shell sweeps up mass as it expands. In the case
of a shell expanding into a vacuum, $A=\frac{3}{2}$, $B=\frac{1}{4}$. In the shells
studied in Paper I and in this paper, the expansion velocity $V$ of the shell is small
compared with $c_{s}$ and so the first two terms in Equation \ref{eqn:disp_rel} are
small.\\
\indent In \cite{2010arXiv1005.4399W}, we derived an alternative expression for the dispersion relation which drops the assumption that the shell is thin and takes into account the effect of the external pressure, $P_{\rm EXT}$
\begin{eqnarray}\nonumber\label{omega_epsilon}
\omega_\epsilon\!&\!=\!&\!-\frac{V\OO}{2\epsilon R\OO}
+\left\{\left(\frac{V\OO}{2\epsilon R\OO}\right)^2\right.\\\nonumber
&&\hspace{1.35cm}+\,\frac{3\pi G\Sigma\OO}{4\epsilon}\left[\frac{r\OO^2\,\cos^{-1}
\left(z\OO / r\OO\right)}{\left(r\OO^2-z\OO^2\right)^{3/2}}\,
-\,\frac{z\OO}{\left(r\OO^2-z\OO^2\right)}\right]\\\nonumber
&&\hspace{1.35cm}+\frac{10P_{_{\rm EXT}}c_{_{\rm S}}^2 l^2}{3\pi^2\epsilon
R\OO^2(2P_{_{\rm EXT}}+\pi G\Sigma\OO^2)}\\
&&\hspace{1.35cm}\left.-\,\frac{5 c_{_{\rm S}}^2 l^2}{2\pi^2\epsilon
R\OO^2}\right\}^{1/2}\, ,
\label{eqn:disp_rel_pagi}
\end{eqnarray}
where $z_{0}$ and $r_{0}$ are respectively the vertical half--thickness and azimuthal radius of a perturbation and $\epsilon$ is a normalisation parameter, described in \cite{2010arXiv1005.4399W}. In order to be useful in the study of star formation, these dispersion relations must be converted into fragment mass functions.\\
\indent Beginning from the thin--shell model, \cite{1994MNRAS.268..291W} neglect the shell expansion terms and consider only the
interplay between gravity and gas pressure within the shell and the masses of fragments formed. They consider
perturbations of radius $r$, compute the acceleration $g$ of material inside them, and write the timescale on which such a fragment
condenses out of the shell as
\begin{eqnarray}
t_{\rm
frag}\approx\left(\frac{r}{g}\right)^{1/2}\approx\left[\frac{G\Sigma_{0}}{r}-\left(\frac{c_{s}}{r}\right)^{2}\right]^{-1/2}.
\label{eqn:t_frag}
\end{eqnarray}
\cite{1994MNRAS.268..291W} then find the value of $r$ that minimises Equation
\ref{eqn:t_frag} and take the mass corresponding to this fragment size as
representative of the mass function. \cite{2007MNRAS.375.1291D} simulate the early stages of
the fragmentation of a shell driven by an expanding HII region and find reasonable
agreement with \cite{1994MNRAS.268..291W} in terms of the time and shell radius at
which fragmentation begins and the mean masses of fragments formed.\\
\indent \cite{2001A&A...374..746W} introduce a more sophisticated analysis in an attempt to derive the mass spectrum of fragments, but their derivation is not correct, since there is an error in their Equation 52. Although we import some of their results, we provide here a different semi--analytic derivation.\\
\indent We seek an expression for the number of fragments $dN$ existing at a time $t$ in the range of masses $\left[m_{\rm f}, m_{\rm f}+{\rm d}m_{\rm f}\right]$ and we make the assumption that the right--hand side of this expression can be separated into two parts; the fragmentation integral (now written as a function of mass) $I_{\rm f}(m_{\rm f},t)$), which expresses how much perturbations of a given mass have grown at time $t$, and a geometrical function $\mathcal{G}(m_{\rm f})$d$m_{\rm f}$ which encodes how many fragments of a given mass will fit on the shell. We may then write
\begin{eqnarray}
dN(m_{\rm f},t)=I_{\rm f}(m_{\rm f},t)\mathcal{G}(m_{\rm f})\,{\rm d}m_{\rm f}.
\label{eqn:dNdl}
\end{eqnarray}
The fragmentation integral is simply defined as 
\begin{eqnarray}
I_{\rm f}(m_{\rm f},t)=\int_{t_{0}}^{t}\omega(m_{\rm f},t^{\prime}){\rm d}t^{\prime} ,
\label{eqn:if}
\end{eqnarray}
where $\omega(m_{\rm f},t^{\prime})$ is the instantaneous growth rate of the mode corresponding to $m_{\rm f}$ and $t_{0}$ is the time at which fragmentation begins.\\
\indent The derivation of the function $\mathcal{G}(m_{\rm f})$d$m_{\rm f}$ is the point at which we depart from the methodology of \cite{2001A&A...374..746W}. It is not clear to us that an analytic expression for $\mathcal{G}(m_{\rm f})$d$m_{\rm f}$ may be found since, on the reasonable assumption that fragments may not overlap each other, the number of objects of any given wavenumber that may be accommodated by the shell depends on how much space has already been consumed by other fragments. We therefore adopted a Monte Carlo approach in an attempt to derive a semi--analytic approximation to $\mathcal{G}(m_{\rm f})$d$m_{\rm f}$.\\
\indent We assume that fragments are circles. We took a unit square (the shape of the area to be filled is largely irrelevant, since circles do not tesselate) and populated it with circles chosen uniformly in $\left[m_{\rm min},m_{\rm max}\right]$, $m_{\rm f}$ being the mass of a circle and proportional to its area. We forbade overlapping and continued until the square was covered up to a factor $f$ by circles. We then constructed histograms of the population of circles. In Figure \ref{fig:circles} we show the result of one such experiment in which $f=0.8$ and $m_{\min},m_{\rm max}$ are 1.26$\times10^{-5}$ and 7.9$\times10^{-1}$ respectively.
In Figure \ref{fig:circles_mf}, we show the power laws resulting from hundreds of such experiments in which we varied $f$. Varying the covering factor affects the low--mass end of the mass spectrum, which turns over at small masses due to underrepresentation of small circles. Achieving higher covering factors requires more small circles and pushes the turnover towards smaller masses. At the high--mass end of the mass function, the mass function becomes noisy. However, in between these limits, the slope of the function is very robust. We find that the population of circles can be well approximated by d$N\propto m_{\rm f}^{-2.0}$d$m_{\rm f}$.\\
\indent Armed with this result we may write
\begin{eqnarray}
dN(m_{\rm f},t)=I_{\rm f}(m_{\rm f} ,t)m_{\rm f}^{-2.0}{\rm d}m_{\rm f}.
\label{eqn:dNdl2}
\end{eqnarray}
\indent An analytic expression for the fragmentation integral itself can be derived for the infinitesimally thin shell and is given in Appendix A in a dimensionless form that can be applied to any thin shell. However, it is somewhat cumbersome and we compute fragmentation integrals from numerically integrating Equation \ref{eqn:if} using analytic wavenumber growth rates for the thin--shell and PAGI models and compare the result to that computed from our SPH simulations.
\section{Numerical methods}
\indent We make use of the same SPH code used in Paper I, a variant of that described in \cite{1990nmns.work..269B} but more recently updated and described in \cite{1995MNRAS.277..362B}. The fluid equations are solved using the SPH technique implementing the standard artificial viscosity prescription described in \cite{1992ARA&A..30..543M}, with $\alpha=1$, $\beta=2$. The self gravity of the gas is included using a binary tree. Crucially for these calculation, the code allows very high density regions to be replaced by point--mass sink particles, as described in \cite{1995MNRAS.277..362B}.\\
\indent Once the density of a particle exceeds a threshold, set to $10^{-19}$ g cm$^{-3}$ in these simulations, it and its $\approx 50$ neighbours are considered candidates for sink creation. In order for a dense particle and its neighbours to be replaced by a sink, the group of particles must (a) constitute more than a thermal Jeans mass; (b) be contracting; (c) be bound. Gas particles may subsequently be accreted by sink particles if they pass within the sink particle's accretion radius and (a) are bound to the sink; (b) are more bound to that sink than to any other sink; (c) do not have enough angular momentum to achieve a circular orbit around the sink. The mass resolutions in the medium-- and high--pressure simulations presented here are 1.67M$_{\odot}$ and 0.4M$_{\odot}$ respectively. We use a sink accretion radius of 0.05pc which ensures that the accretion flows at the accretion radius are always supersonic, so that accretion of gas particles does not produce spurious drops in pressure. We repeated part of the medium--pressure calculation using an accretion radius of 0.025pc, but found that this made no difference to the results. The code also allows for corrections to be made to the pressure and density gradients in the neighbourhood of sink particles to compensate for the fact that there are few or no gas particles inside the accretion radius, but we found that they had no effect on our results. In these calculations, the gas flows are non--turbulent and thus rather quiet and smooth, and hence comparatively easy to model.\\
\indent To achieve greater accuracy, the sink particles are not included in the binary tree and all gravitational forces involving them are computed by direct summation. To prevent sink particles in very close proximity to each other acquiring very short timesteps and stalling the simulation, we allowed point masses to merge if they passed within 0.05pc of each other and were mutually bound. During the course of the simulations, there are a few tens of such mergers, too few to strongly influence the mass functions produced.\\
\indent We use the same spherical harmonic decomposition technique discussed in Paper I to analyse the growth of perturbations in the shell surface density, which we compare to the thin shell model. The thin--shell model uses linearized equations and is therefore only guaranteed to be valid in the linear phase of fragmentation, where surface density perturbations are less than or comparable to the mean surface density. In this work we are interested in the transition of the shell into the non--linear regime and the contraction of such perturbations into bound objects. Apart from this phase of fragmentation being non--linear, there also comes a point for each contracting fragment when its wavelength becomes comparable to the thickness of the shell and fragmentation transitions from a two--dimensional to a three--dimensional process. In addition, the wavelengths and wavenumbers of fragments changes as they collapse, and the surface density associated with the fragment grows, so that the association of a given wavenumber with a given mass may become invalid. For these reasons, it is not clear that the thin--shell or PAGI analysis may be extended into this regime of the shell's evolution and we therefore turn to a numerical study.\\
\indent We use two different techniques to follow the fragmentation process beyond the linear phase. Once fragmentation has become strongly non--linear and bound objects are collapsing on their local freefall times, we make use of SPH sink particles \citep{1995MNRAS.277..362B}. However, in the regime between the linear and strongly non--linear phases of the shell evolution, before any fragments have become gravitationally bound, we require a robust clump--finding method.\\
\indent Three dimensional clump--finding in complex density fields is a difficult task and several numerical methods have been developed to facilitate it, e.g. the CLUMPFIND \citep{1994ApJ...428..693W} algorithm. In a particle--based code such as SPH, progress can be made in identifying perturbations in the density field by identifying contiguous regions of fluid whose density exceeds some threshold. Constructing a mass function is problematic however, since the chosen density threshold effectively selects the masses of the fragments. This problem can be solved by insisting that fragments must be bound. By selecting the densest particles as seeds for clumps and continually adding adjacent particles until the total energy of the composite object becomes positive, it is possible to identify a population of bound objects.\\
\indent This method was used by \cite{2007MNRAS.375.1291D}, but is of little use here, since it is of no help in the mildly non--linear regime. We find that a given object goes very rapidly from becoming bound to forming a sink particle, so that this method reveals little that cannot be learned from the sink particles themselves. We therefore make use of the clump--finding method of \cite{2009MNRAS.396..830S} which identifies structure using peaks in the gravitational potential field, instead of in density. There are two advantages to such a method. First, the gravitational potential distribution in the cloud is considerably smoother than the density distribution, since density fluctuations that do not carry sufficient mass cannot contribute significantly to the potential field. Second, the strength of the gravitational potential determines whether a clump will collapse and how mass will flow. The structures identified by this algorithm are called p-cores to distinguish them from conventional density cores.\\
\begin{figure}
\includegraphics[width=0.45\textwidth]{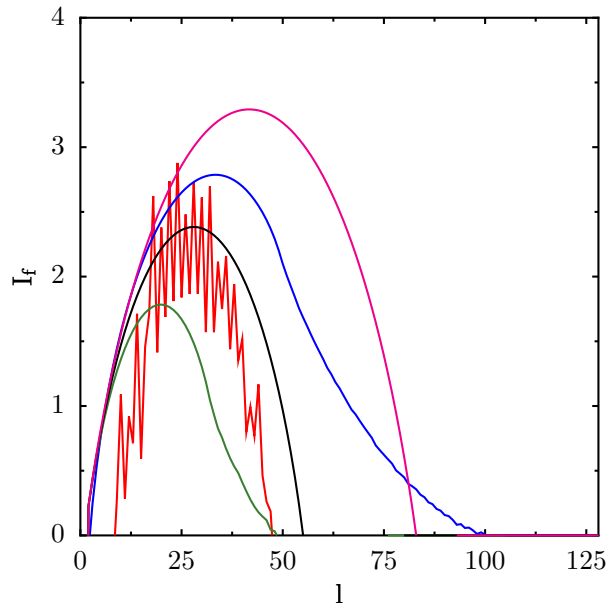}
\caption{Comparison of the PAGI low--pressure ($1\times10^{-17}$ dyne cm$^{-2}$, green), PAGI medium--pressure ($1\times10^{-13}$ dyne cm$^{-2}$, black), PAGI high--pressure ($5\times10^{-13}$ dyne cm$^{-2}$, magenta), thin--shell (blue) and numerical (red) medium--pressure fragmentation integrals computed up to a time of 8.80 Myr.}
\label{hello}
\end{figure}
\begin{figure}
\includegraphics[width=0.45\textwidth]{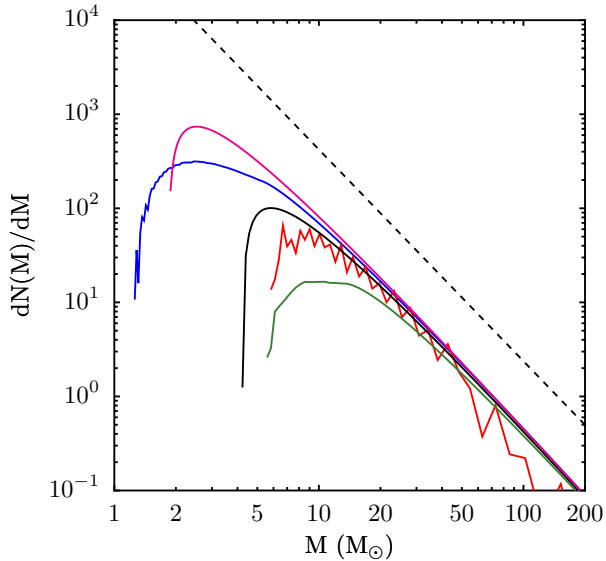}
\caption{Comparison of the PAGI low--pressure ($1\times10^{-17}$ dyne cm$^{-2}$, green), PAGI medium--pressure ($1\times10^{-13}$ dyne cm$^{-2}$, black), PAGI high--pressure ($5\times10^{-13}$ dyne cm$^{-2}$, magenta), thin--shell medium--pressure (blue) and numerical medium--pressure (red) mass functions, derived from Equation \ref{eqn:dNdl2}, computed up to a time of 8.80 Myr. The dashed line has a logarithmic slope of -2.25.}
\label{fig:mf_real_theory}
\end{figure}
\indent The potential--clumpfinding algorithm works in a similar manner to CLUMPFIND and can be applied directly to the SPH data. The SPH particle with the deepest gravitational potential forms the head of a clump, then the particle with the next deepest potential is either assigned to the same clump if it is an SPH neighbour of the head particle, or forms a new clump if not, and so on. Clumps are defined down to either a minimum positive potential, or the lowest contour which they share with a neighbouring clump. Unlike the traditional CLUMPFIND algorithm, we use contour levels primarily to define the level at which potential clumps join, rather than to distinguish clumps from noise, and so our contours are numerous and finely spaced. This has the effect of subtracting the background potential. This is necessary as gravity is a long range force, and so is affected by both the mass inside the p--core and surrounding it; hence we must remove the background to obtain the net effect on the mass within. P--cores, therefore, represent the local maximum above the surrounding background.
\begin{figure}
\includegraphics[width=0.45\textwidth]{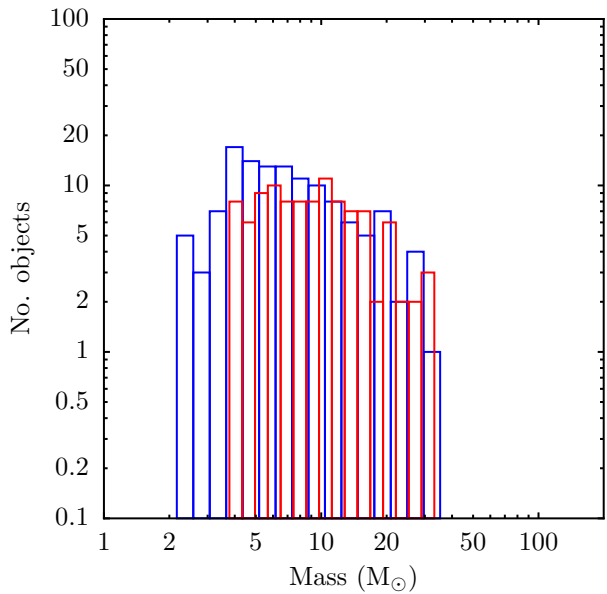}
\caption{Comparison of clump mass functions generated by imposing an artificial surface density profile computed from the fragmentation integral of a given SPH dump on a uniform shell (red) with that computed from analysing the same dump directly (blue).}
\label{fig:fake_mf}
\end{figure}
\section{Simulations}
In Paper I, we presented the results of simulations of expanding momentum--driven shells with two different boundary conditions. The shells had a mass of $2\times10^{4}$M$_{\odot}$, an initial radius of 10 pc and an initial velocity of 2.1 kms$^{-1}$, such that they expand to a maximum radius of $\approx23$pc before beginning to contract. In the simulation with free boundaries, we observed that the shell thickened considerably during its expansion and that this suppressed the growth of fragments at large wavenumbers. We repeated the simulation applying constant and equal pressures of $1.0\times10^{-13}$ dyne cm$^{-2}$ to the inner and outer faces of the shell such that its thickness remained approximately constant and found that this resulted in much better agreement with the thin shell model. However, in Paper I, we followed the shell evolution only in the linear regime in which the thin--shell model is most likely to be applicable. In this paper, we allow the simulations of Paper I to run further, into the non--linear regime, using only the SPH simulations and not the AMR simulations, since both sink formation and clumpfinding are much more difficult in the latter. We find that the fragmentation process in the non--pressure confined simulation from Paper I proceeds too slowly to form a significant number of sink particles before the shell has contracted back to its original size and we do not examine it further.\\
\indent In Paper II, we considered the effects of higher pressures and found that they result in additional shorter wavelengths becoming unstable, which should produce more low--mass fragments. In this work, we concentrate on the pressure--confined model from Paper I (which we refer to as the medium--pressure model), and we also study the mass function produced by the same shell, but with a confining pressure of $5.0\times10^{-13}$ dyne cm$^{-2}$, which we also studied in Paper II and will refer to as the high--pressure model.\\
\section{Results}
\subsection{Spectral analysis vs. clumpfinding}
In Figure \ref{hello} we plot the fragmentation integrals computed for the medium--pressure run from the thin--shell model, for three different pressures using the PAGI model, and that derived from the spherical--harmonic analysis of the medium--pressure SPH simulation, all computed at a time of 8.80Myr. The analytic fragmentation integrals are obtained by integrating Equation \ref{eqn:if}, using Equation \ref{eqn:disp_rel} to compute $\omega$ in the thin--shell case, and Equation \ref{eqn:disp_rel_pagi} to compute $\omega_{\epsilon}$ in the PAGI case. The numerical fragmentation integral clearly agrees much better with the relevant PAGI model than with the thin--shell model. Importing the results from Section 2, we then compute mass functions from the fragmentation integrals, shown in Figure \ref{fig:mf_real_theory}. We again see that the mass function computed from the PAGI model is in much better agreement with that computed from the SPH medium--pressure calculation than is the thin--shell mass function. We find that, at high masses, the theoretical mass functions are power laws with slopes of -2.25, in good agreement with the canonical Salpeter slope of -2.35.\\
\indent Before proceeding with clumpfinding, we sought to establish a connection between the two--dimensional spectral analysis technique and the intrinsically three--dimensional concept of a clump defined by its gravitational potential. We selected an output file from the early stages of the medium--pressure simulation and computed the corresponding fragmentation integral. We then used the integral to construct a perturbed surface density profile which we imposed on a uniform shell by iteratively adjusting the SPH particle masses in a manner similar to that described in Paper I. Finally, we computed a clump mass function from the artificial shell and compared it to that computed from the original output file from the medium--pressure simulation. As shown in Figure \ref{fig:fake_mf}, the mass functions generated in this way are in reasonable agreement, which implies that the spectral analysis technique is a legitimate way of describing shell fragmentation, at least in the early linear regime.\\
\subsection{Mass function of clumps}
\begin{figure}
\includegraphics[width=0.45\textwidth]{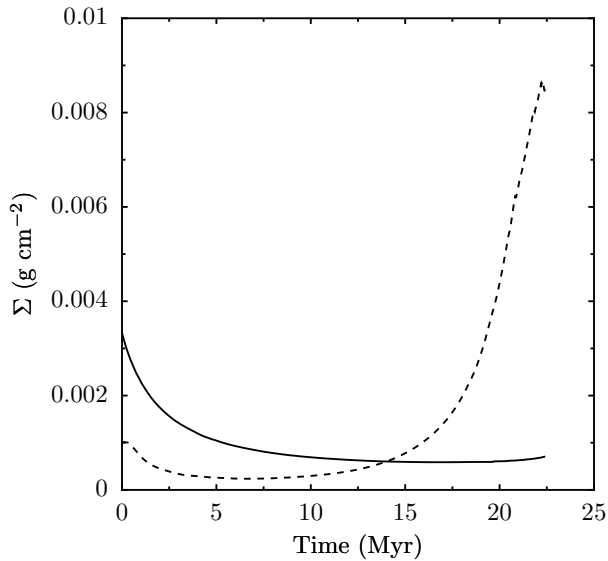}
\caption{Comparison of mean (solid line) and maximum perturbed (dashed line) surface density in the medium--pressure calculation.}
\label{fig:sig_max}
\end{figure}
\begin{figure*}
\subfigure{\includegraphics[width=0.30\textwidth]{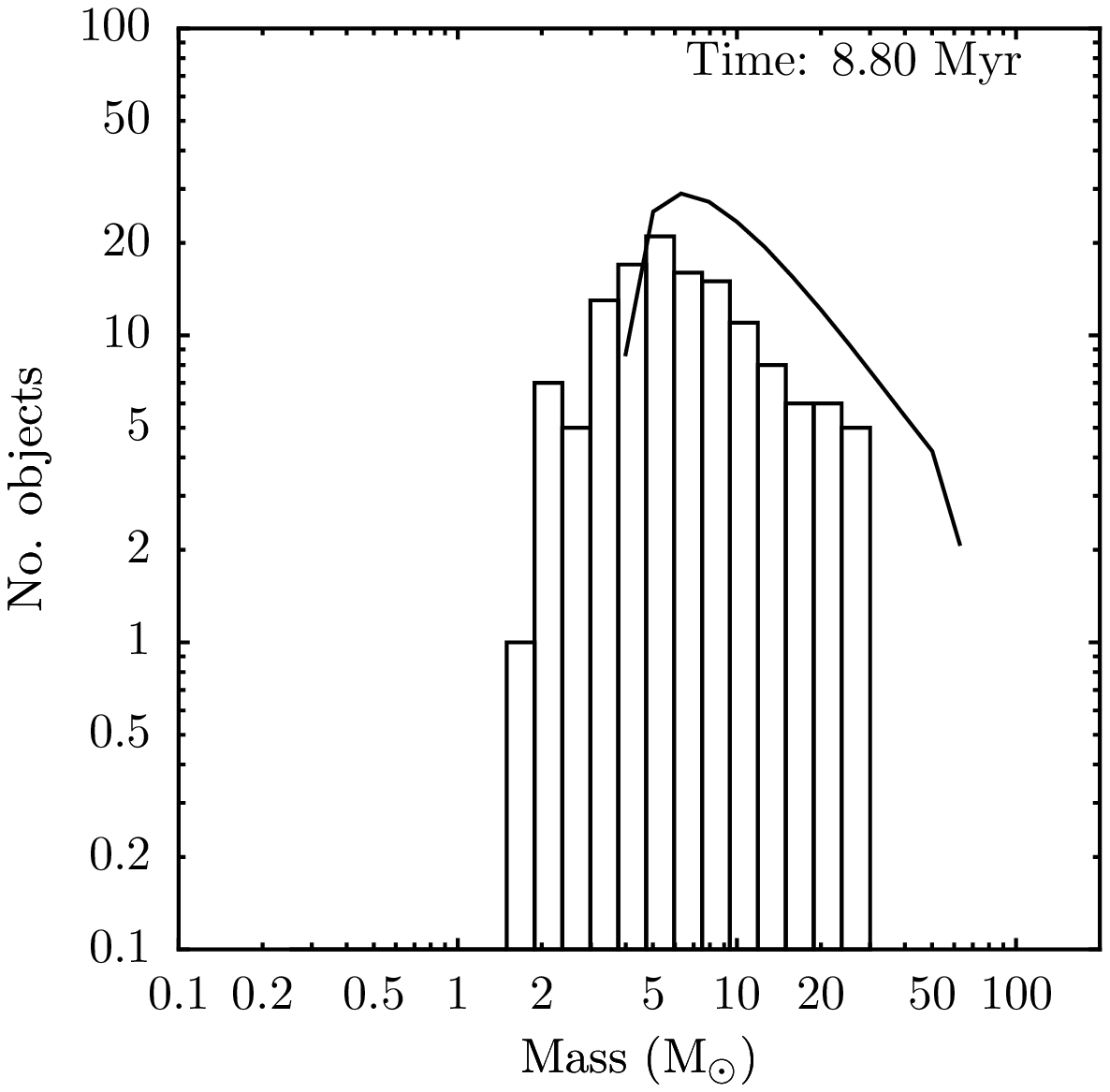}}
\hspace{0.1in}
\subfigure{\includegraphics[width=0.30\textwidth]{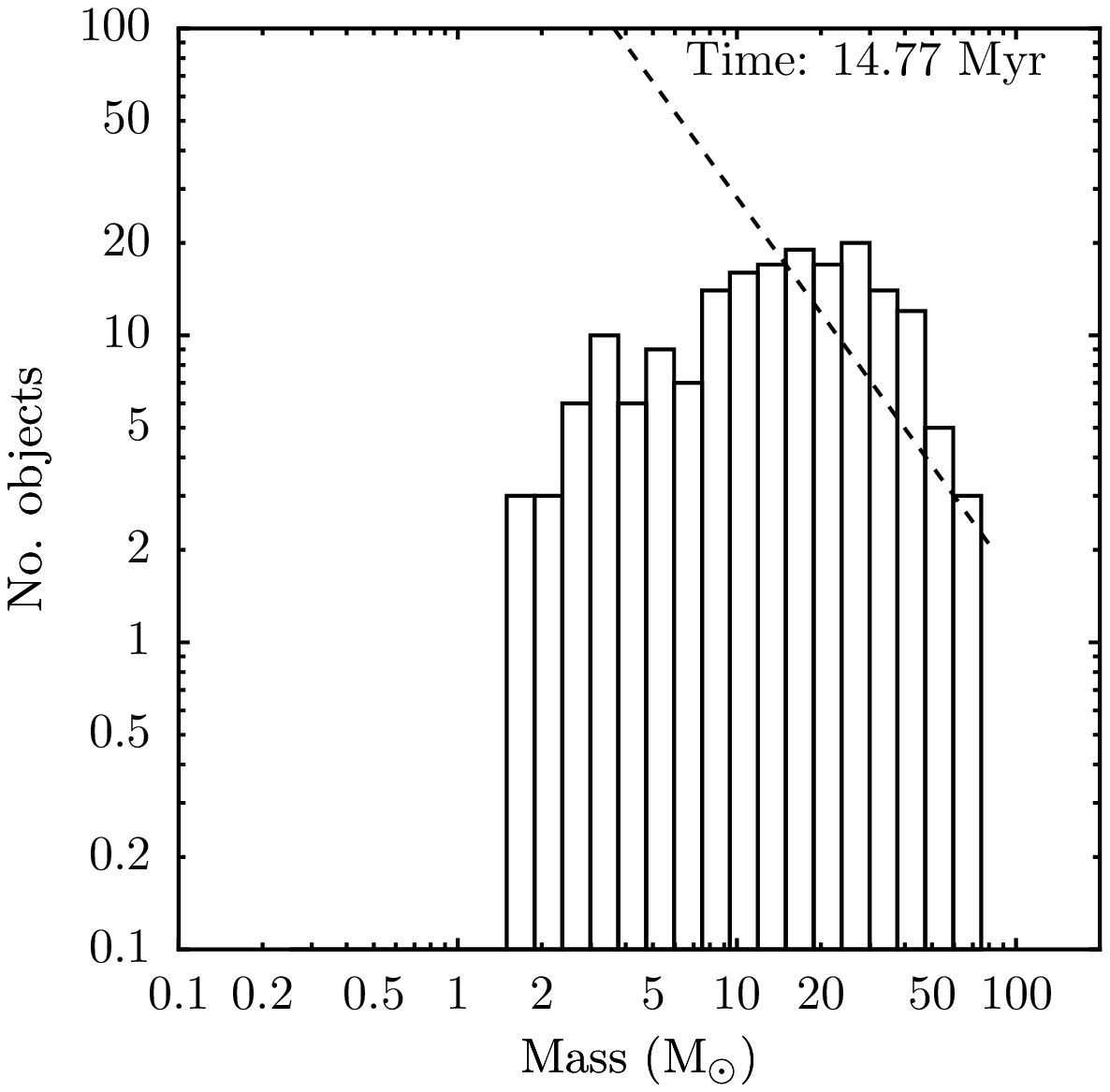}}
\hspace{0.1in}
\subfigure{\includegraphics[width=0.30\textwidth]{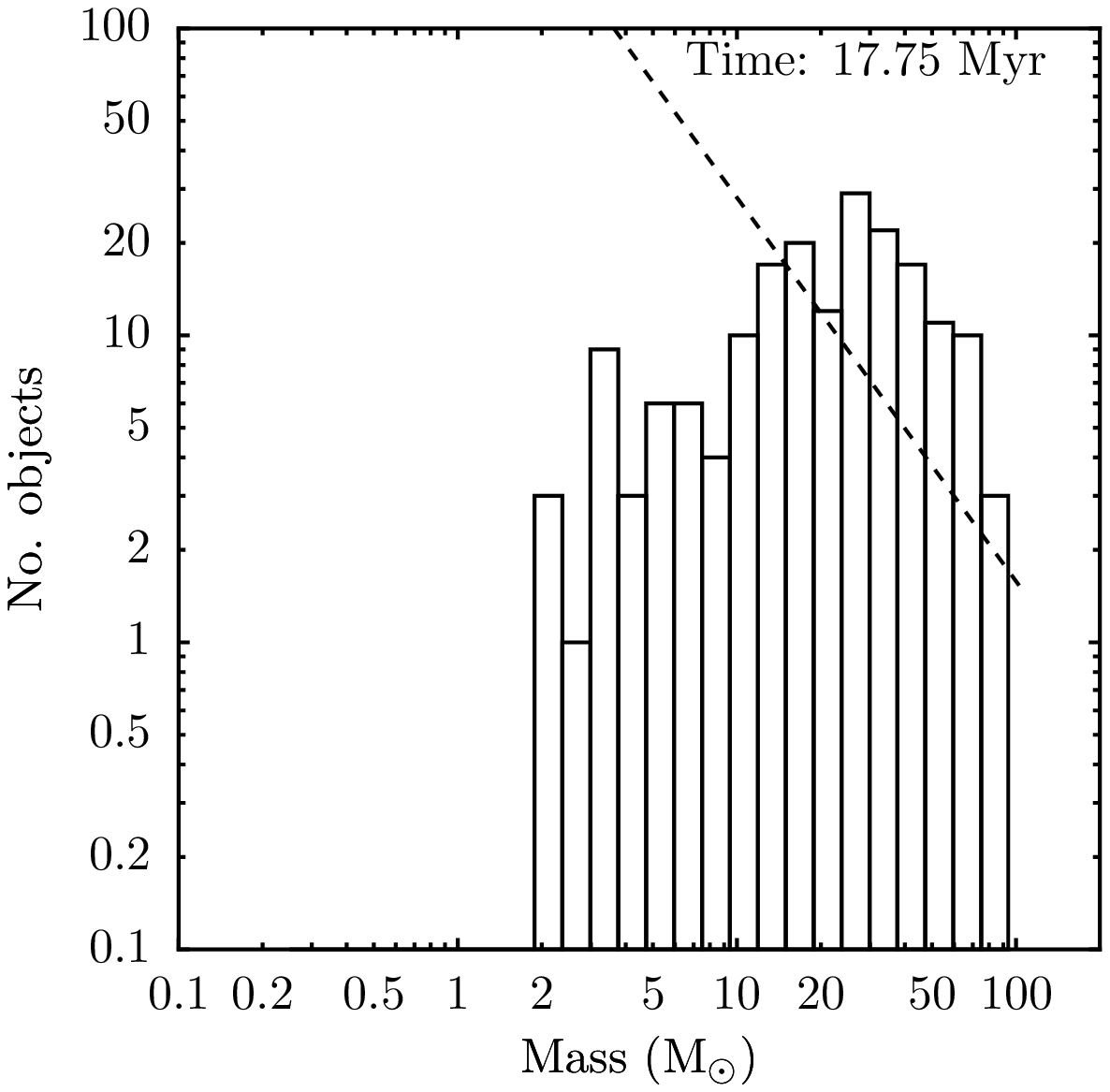}}
\vspace{0.1in}
\subfigure{\includegraphics[width=0.30\textwidth]{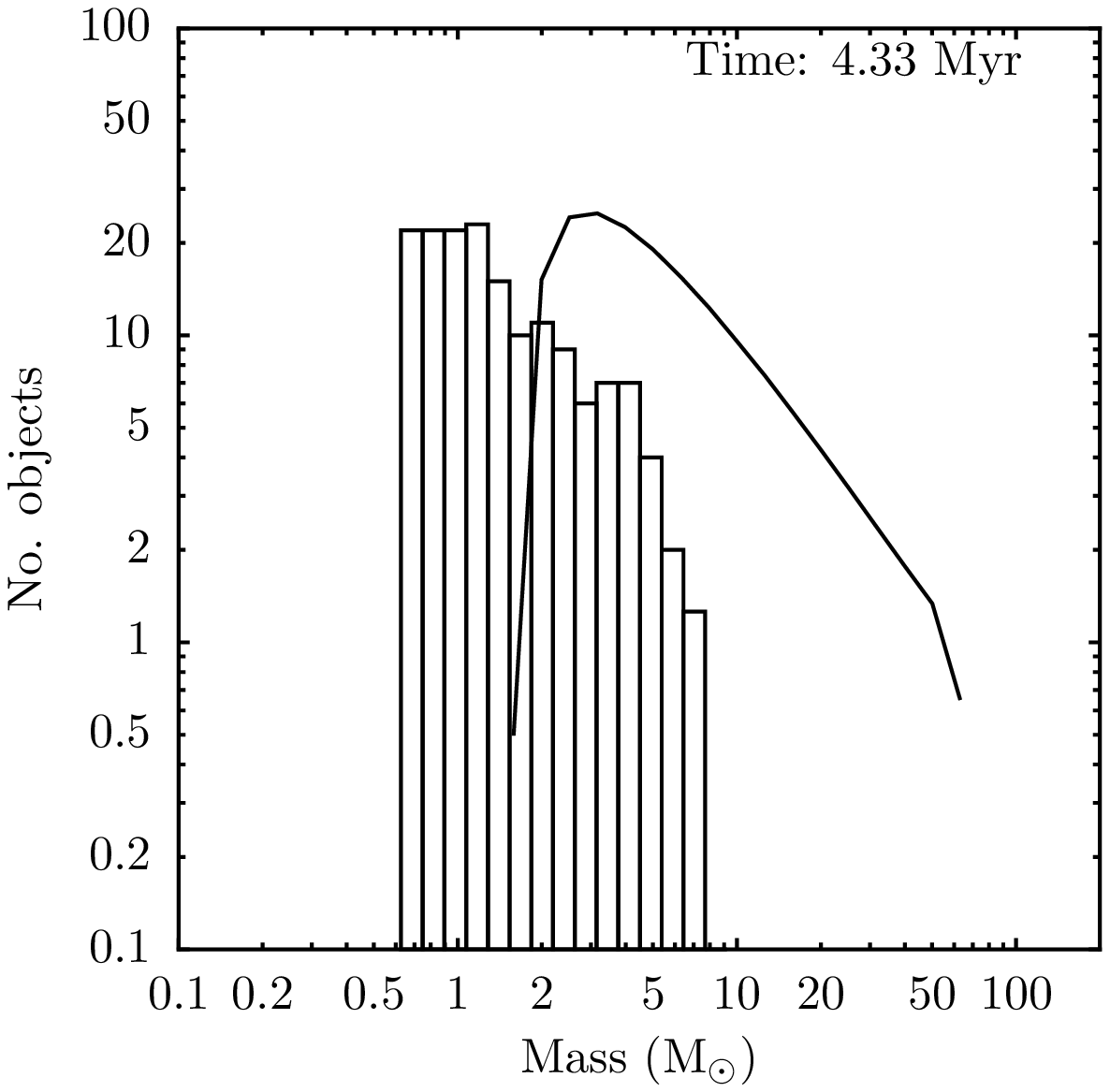}}
\hspace{0.1in}
\subfigure{\includegraphics[width=0.30\textwidth]{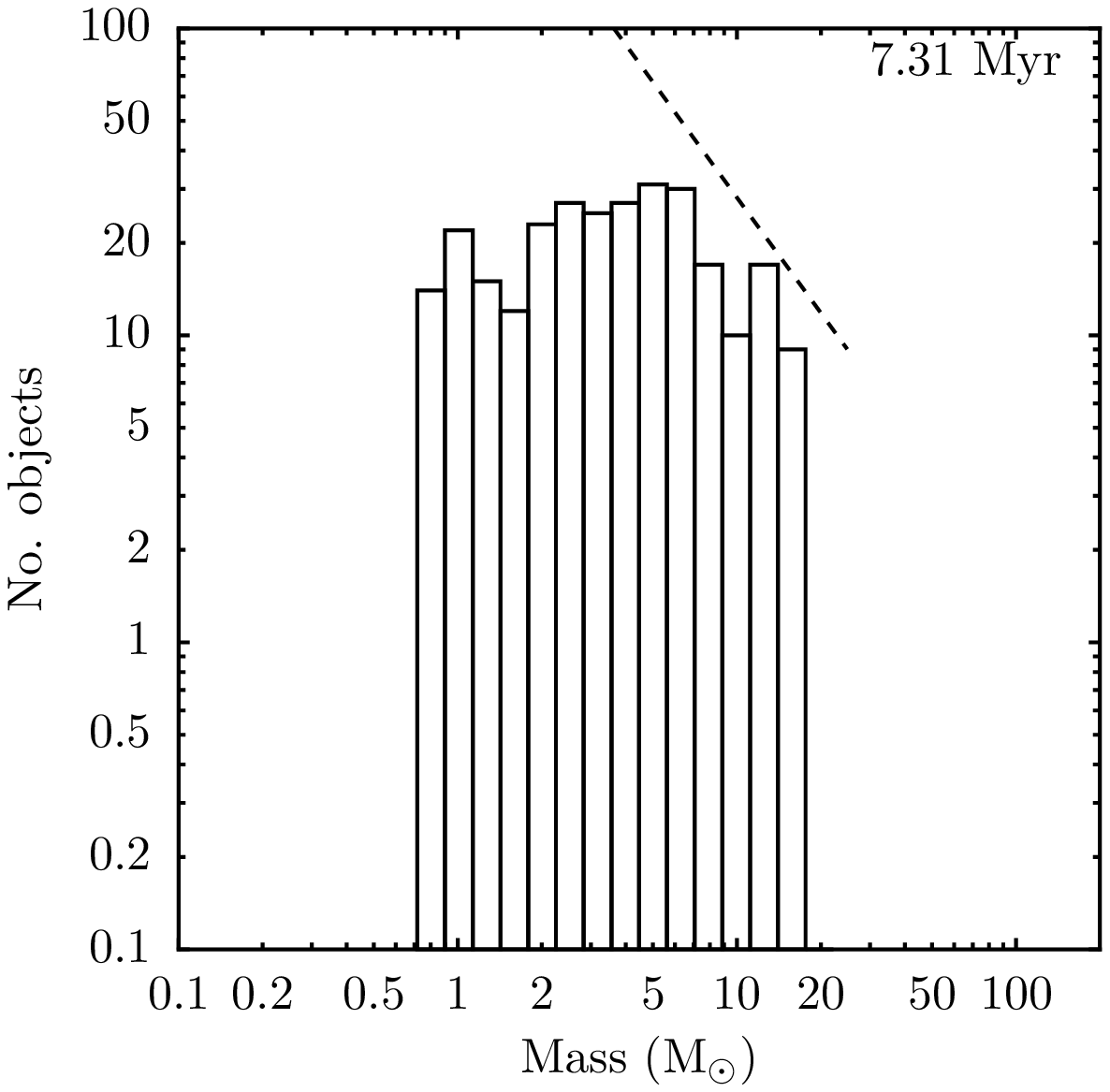}}
\hspace{0.1in}
\subfigure{\includegraphics[width=0.30\textwidth]{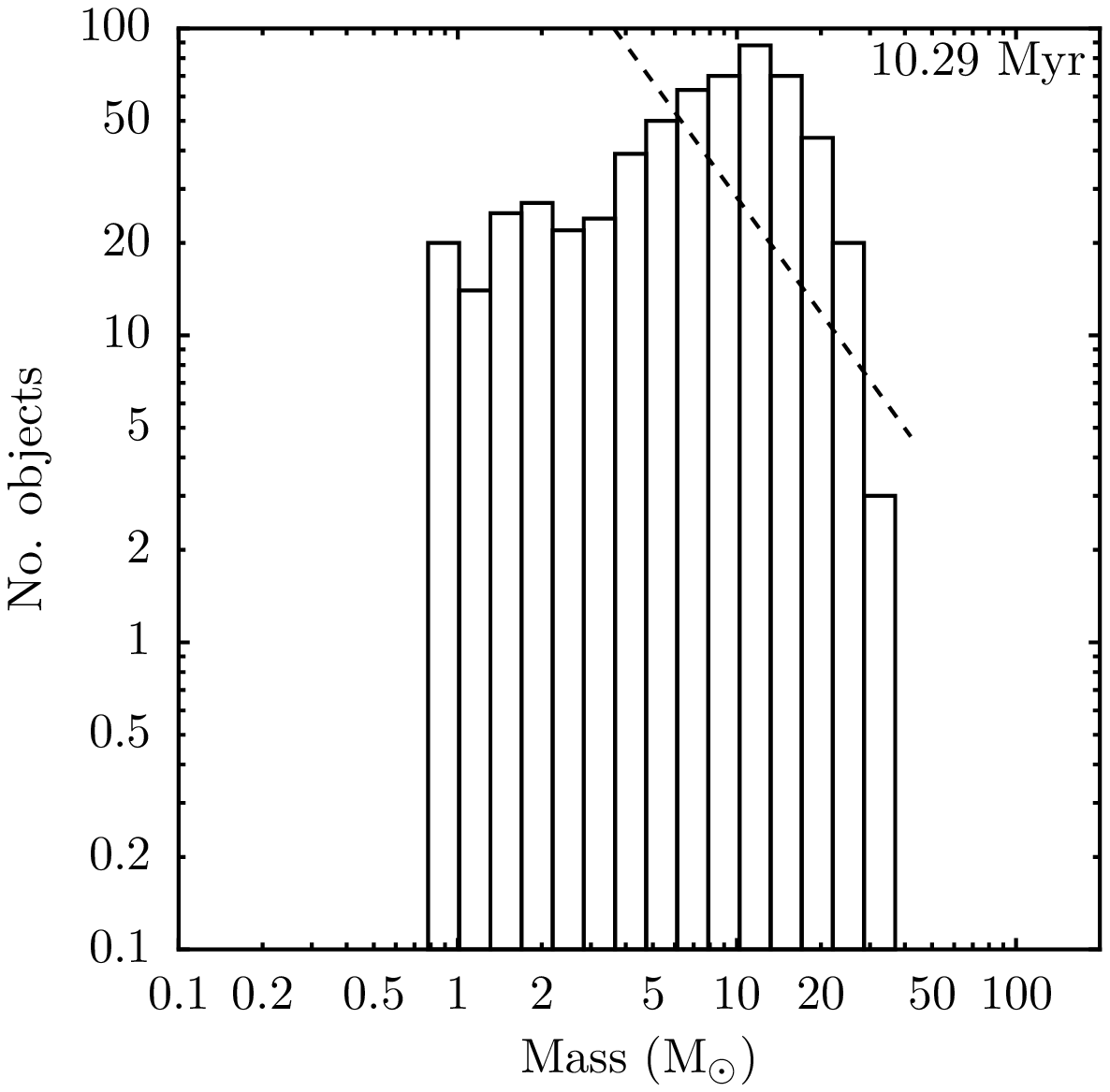}}
\vspace{0.1in}
\caption{Mass functions of clumps identified by the potential--based clump finder in the medium--pressure calculation (top row) and in the high--pressure calculation (bottom row) at three epochs (histograms). Left panels show the mass functions derived from the PAGI model (solid lines) and centre and right panels show power--laws with a logarithmic slope -2.25 (dashed line).}
\label{fig:clump_mf}
\end{figure*}
\begin{figure}
\includegraphics[width=0.45\textwidth]{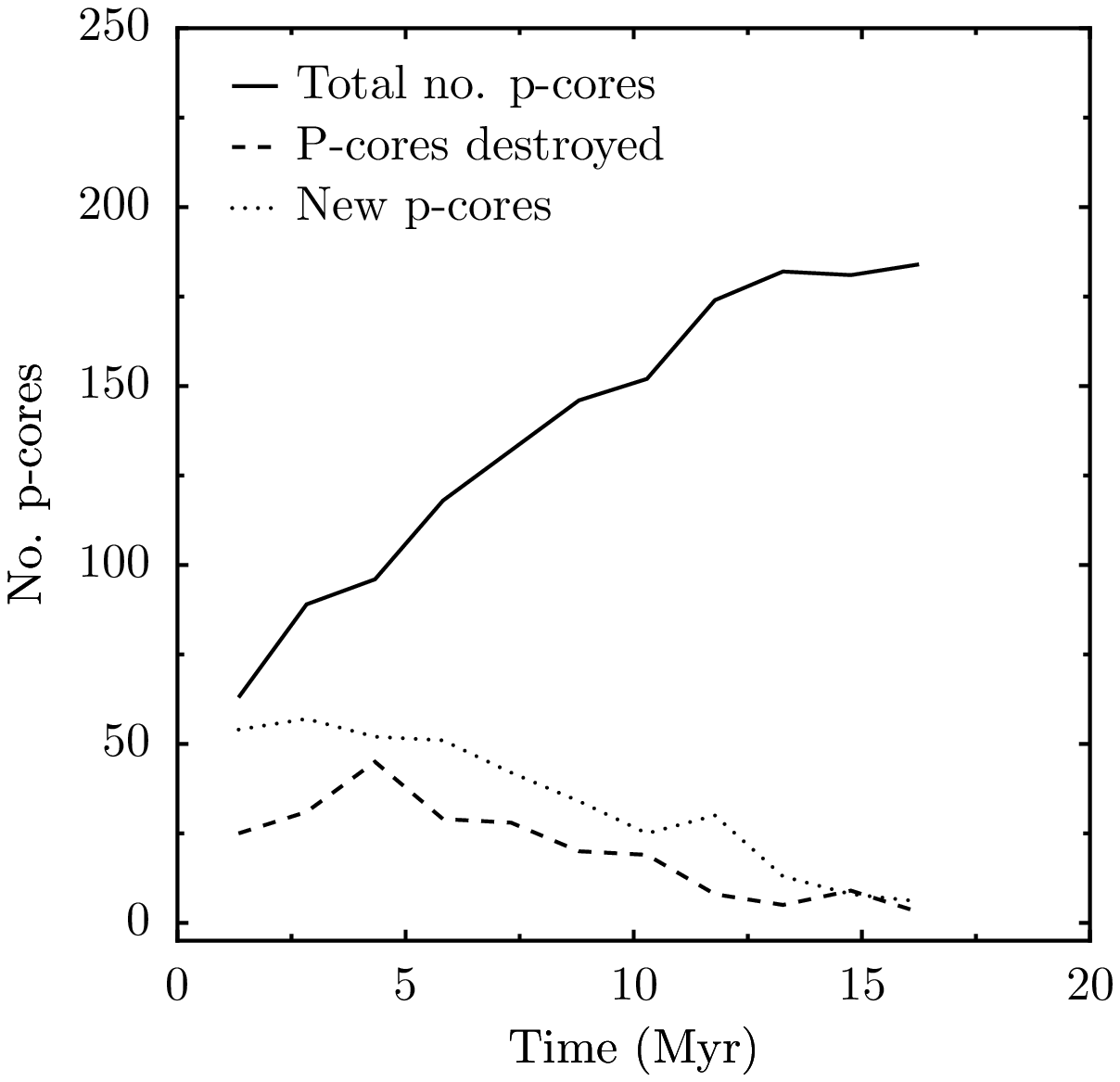}
\caption{Total number of p--cores found by the clumpfinder in the medium--pressure run as a function of time (solid line), number of p--cores destroyed in the previous epoch (dashed line) and number of new p--cores formed in the previous epoch (dotted line). Epochs are 1.5Myr in duration..}
\label{fig:nclumps}
\end{figure}
\begin{figure}
\includegraphics[width=0.45\textwidth]{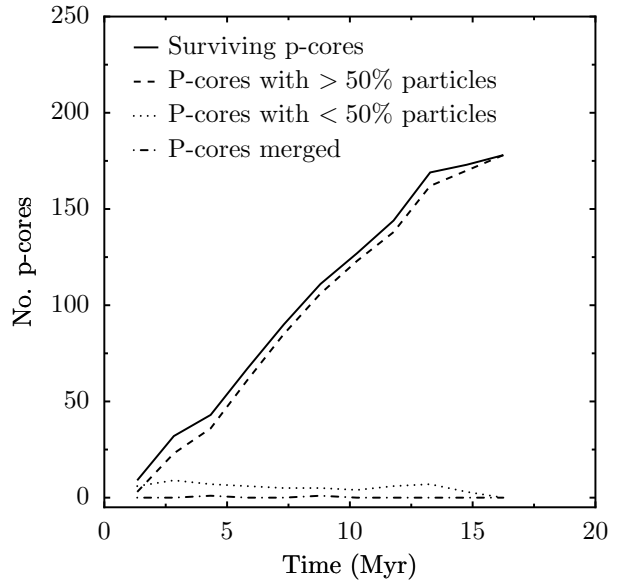}
\caption{Number of p--cores surviving from previous epoch (solid line), number of p--cores retaining more than half of the particles assigned to them in the previous epoch (dashed line), number of p--cores retaining less than half of the particles assigned to them in the previous epoch (dotted line) and number of p--cores involved in mergers in the previous epoch (dash--dotted line), all plotted as functions of time, and all from the medium--pressure run. Epochs are 1.5Myr in duration.}
\label{fig:ninteract}
\end{figure}

In Paper I, we performed tests in which we inserted monochromatic density perturbations into the shell and compared their evolution with that predicted by the thin shell model. We found that the agreement was very good until the point when the perturbation in the surface density became equal to the mean surface density, after which the perturbation growth became non--linear, departing strongly from the model. In Figure \ref{fig:sig_max}, we plot the mean shell surface density (solid line) and the maximum perturbation in the surface density (defined as the maximum surface density minus the mean surface density, dashed line). We see that the time at which the surface density perturbation comes to exceed the mean surface density, and therefore the point at which perturbation growth is expected to become non--linear, occurs at $\approx13$Myr.\\
\indent In Figure \ref{fig:clump_mf} we plot the mass functions of the p--cores detected by the clumpfinding procedure at three epochs in the medium-- and high--pressure calculations, and the slope derived from integration of the analytical fragmentation integral. We find that, in the early stages of the shell's evolution (left panels) while the shell's evolution is still in the linear regime, the mass spectrum of fragments identified by the clump--finder follows has a form very similar to that predicted by the PAGI model, although the p--core mass function appears to be offset somewhat towards lower masses. There are two reasons for this. Firstly, material on the outskirts of a given perturbation is rejected by the potential clump--finding algorithm as being part of the background, so that the algorithm underestimates the masses of fragments. More importantly, the association of perturbations of wavelength $\lambda$ with objects of mass $m_{f}$ requires an assumption about the radius $r$ of the object in terms of the wavelength. We adopt $r=\lambda/2$, but this likely overestimates the masses of fragments as it includes material whose density is little different from that of the background shell.\\
\begin{figure*}
\subfigure{\includegraphics[width=0.30\textwidth]{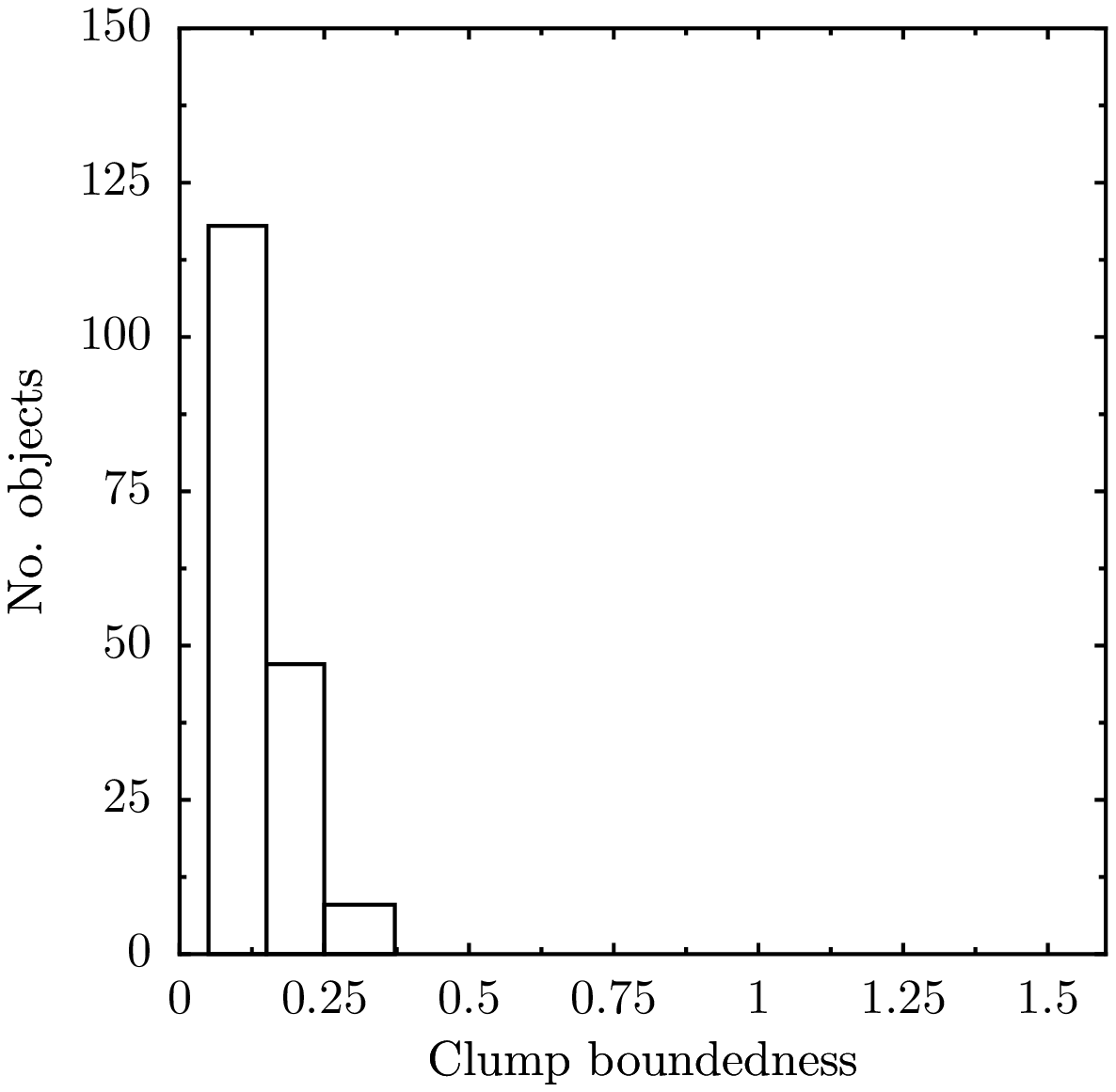}}
\hspace{0.1in}
\subfigure{\includegraphics[width=0.30\textwidth]{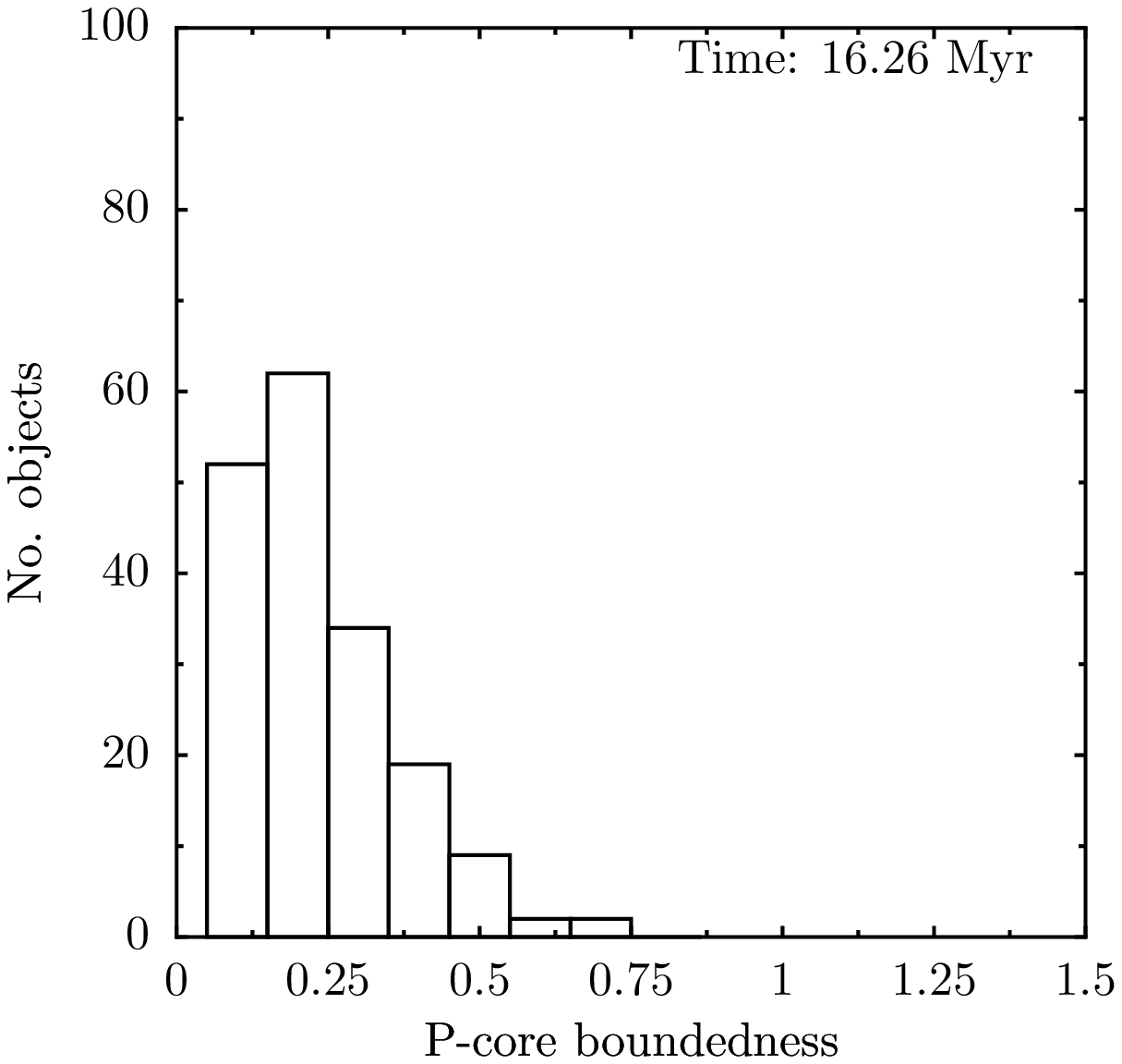}}
\hspace{0.1in}
\subfigure{\includegraphics[width=0.30\textwidth]{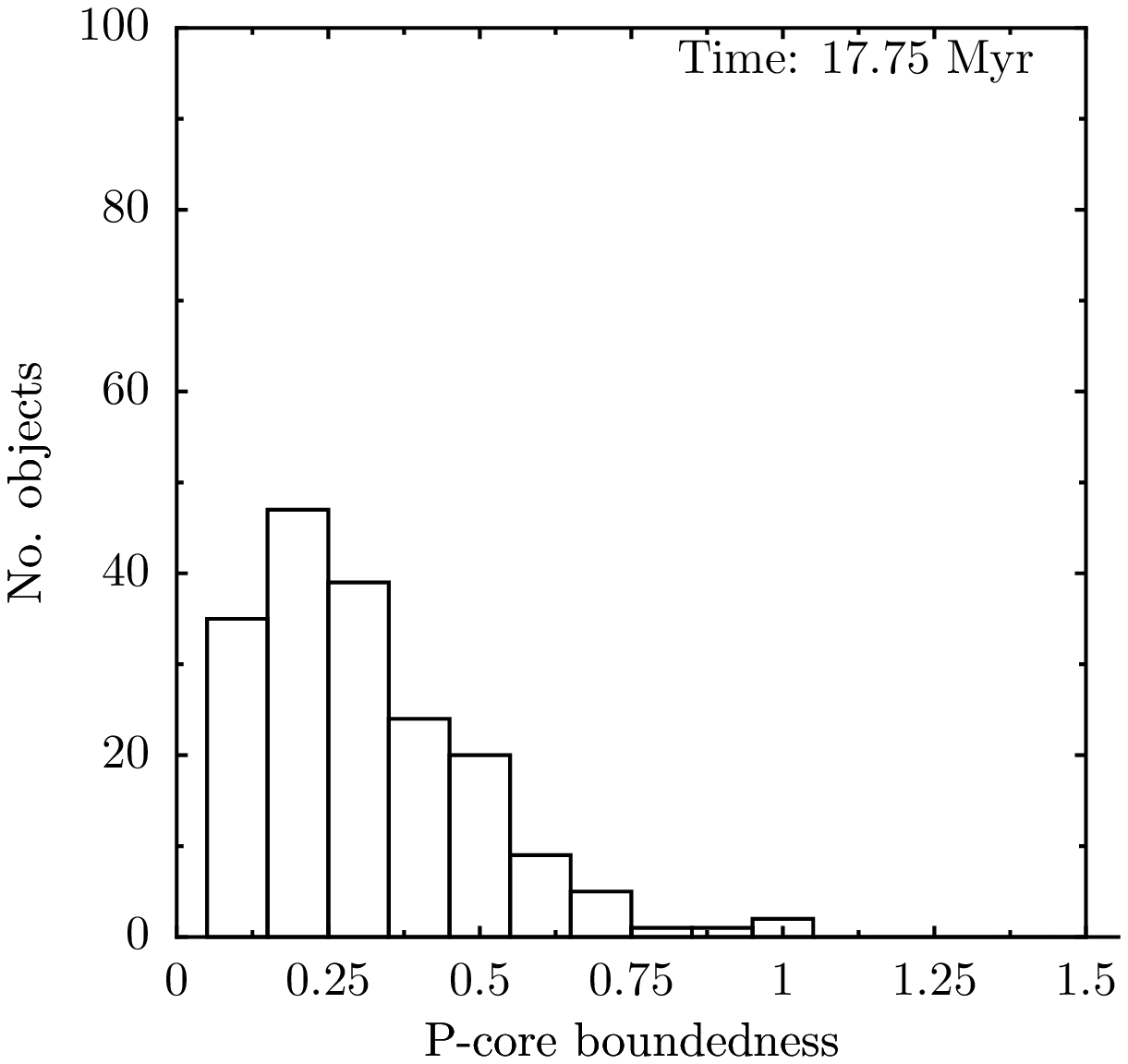}}
\caption{Boundedness (as defined by Equation \ref{Erat}) of fragments detected by the potential clump--finding algorithm at three epochs in the medium--pressure calculation.}
\label{fig:clump_bnd}
\end{figure*}
\indent At later times (but before the formation of the first sink particles) the mass function of the p--cores becomes top--heavy (centre and right panels of Figure \ref{fig:clump_mf}), with a deficit of low--mass  objects, although the slope of the high--mass end of the mass function remains consistent with a power law.\\
\indent In Figure \ref{fig:clump_bnd} we plot the numbers of p--cores found by the clumpfinder (before the formation of any sink particles) against the boundedness of the clumps $E_{\rm rat}$, defined as
\begin{equation}
\label{Erat}
E_{\rm rat}=\frac{|E_{\rm p}|}{E_{\rm therm}+E_{\rm k}},
\end{equation}
where $E_{\rm k}$ is the kinetic energy calculated with respect to the centre of velocity of the p--core, $E_{\rm therm}$ is its thermal energy and $E_{\rm p}$ is the potential energy of the p--core calculated using the relative depth of the potential well once the background has been subtracted. Thus a p-core gravitational potential depth is expressed relative to the potential level in its vicinity. Values of $E_{\rm rat}$ larger than unity indicate p--cores bound with respect to their environment, not merely when considered in isolation. Figures \ref{fig:sig_max}, \ref{fig:clump_mf} and \ref{fig:clump_bnd} show a clear sequence of events in which the evolution of the shell begins to depart from the predictions of the linear PAGI analysis around the time when the surface density perturbations come to exceed the mean shell surface density. The clump mass function then ceases to be a power law and the perturbations begin to contract and become gravitationally bound.\\
\indent A fundamental assumption of the thin--shell model is that all fragments evolve independently of one another in an otherwise unperturbed shell. We used the potential clump--finding code to verify this assumption. The clumpfinding code is able to follow individual p--cores to see how they gain or lose mass and whether they merge or exchange mass with each other as time progresses. To make use of this facility, we arbitrarily divide the first 15Myr of the medium--pressure shell's evolution into ten contiguous 1.5 Myr epochs. In Figure \ref{fig:nclumps} we plot the total number of p--cores extant in each epoch together with the numbers of clumps formed and destroyed in the previous epoch, against time. Initially, the number of p--cores destroyed in the previous epoch is almost equal to the total number of p--cores, implying that the objects detected by the clumpfinder are transient entities. However, as the simulation progresses, the total number of p--cores rises and the number destroyed falls, so that the objects clearly become more longlived. Around 15Myr, the total number of p--cores levels off (and the numbers of new ones formed and of existing ones dissolving declines almost to zero) implying that the shell evolves to contain a stable number of long--lived objects. In Figure \ref{fig:ninteract} we plot the numbers of p--cores in each 1.5Myr epoch which have survived from the previous epoch, the numbers which have retained more, or less, than half of the particles previously assigned to them and the numbers that have merged since the previous epoch. The number of p--cores surviving from the previous epoch rises rapidly as they become longlived, as opposed to ephemeral. We find that, in the vast majority of cases an SPH particle assigned to a given p--core remains assigned to that p--core unless the p--core itself dissolves, and that mergers or fragmentation of cores are very rare. Our findings therefore support the assumption that surface density perturbations do not interact with each other. We also note that the total number of p--cores levels off at a value of $\sim200$. Figure \ref{hello} shows that the most unstable wavenumber at the time when the first p--cores become bound is $\sim25$. If the mass of a fragment is $\pi\lambda^{2}M/16\pi R^{2}$,  the number of fragments with this wavenumber that the shell can support is $4l^{2}/\pi^{2}\approx250$, in reasonable agreement with the total number of fragments detected by the clumpfinder. \\
\subsection{Sink particles and oligarchic accretion}
\begin{figure*}
\subfigure{\includegraphics[width=0.30\textwidth]{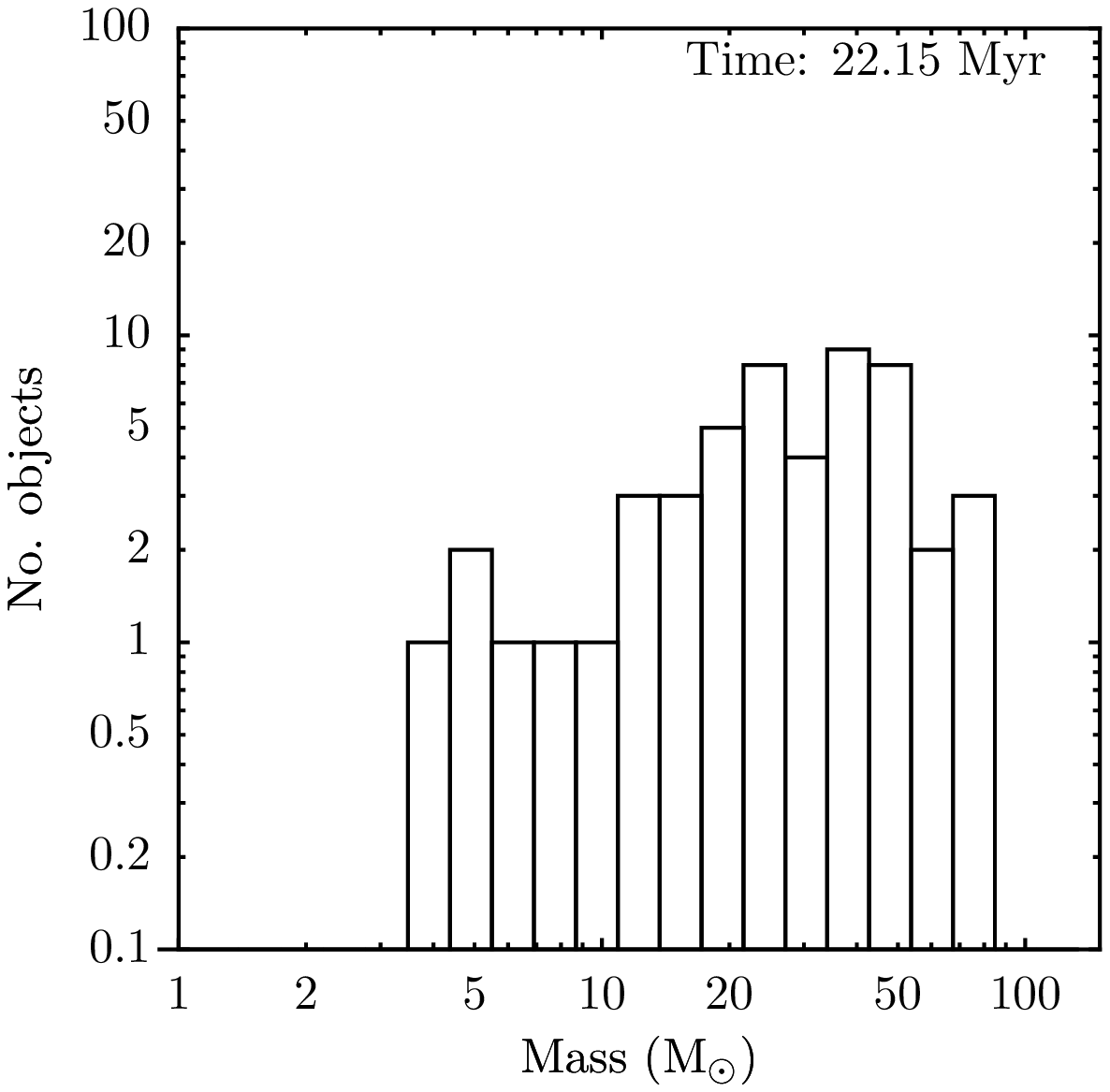}}
\hspace{0.1in}
\subfigure{\includegraphics[width=0.30\textwidth]{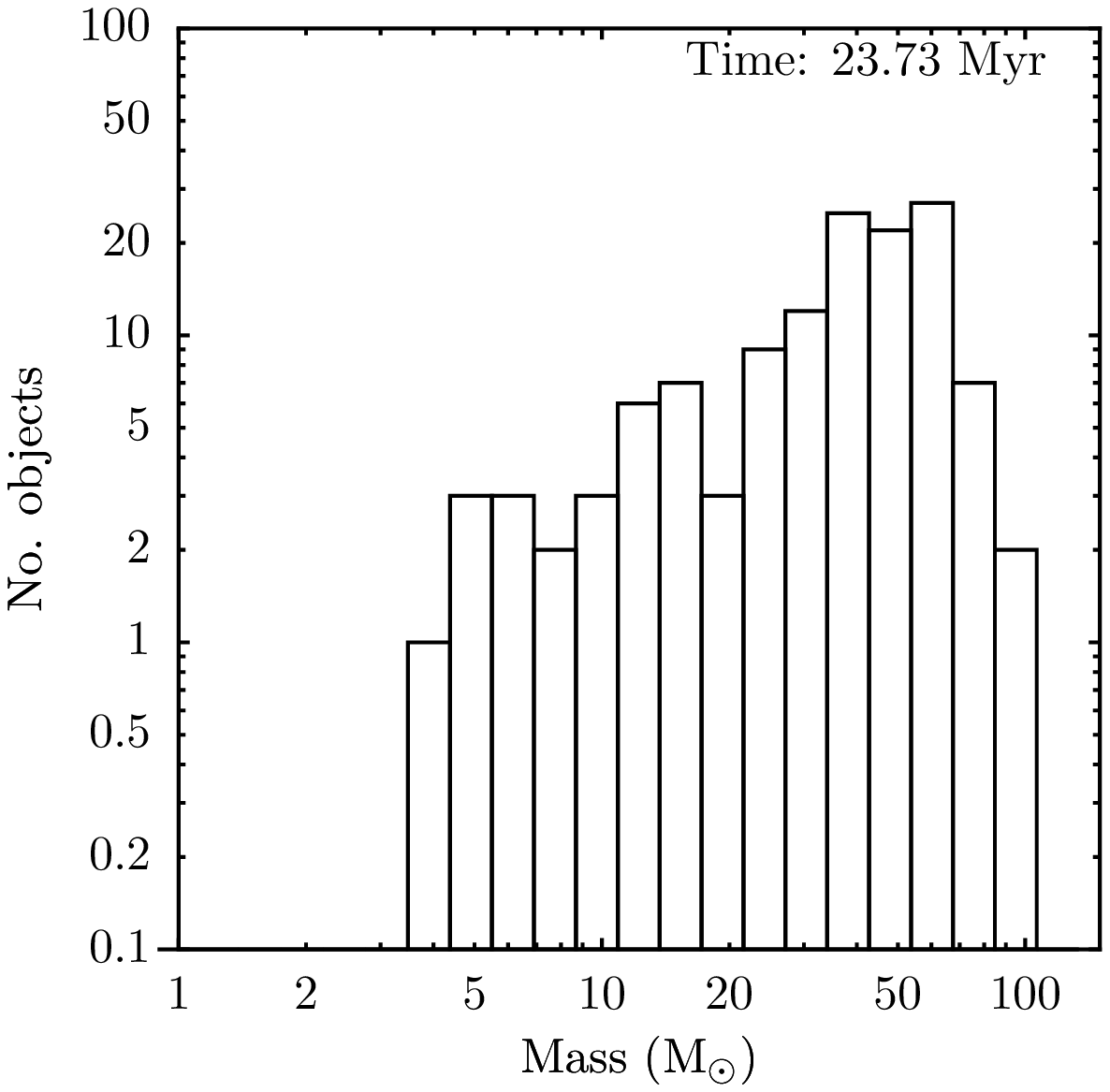}}
\hspace{0.1in}
\subfigure{\includegraphics[width=0.30\textwidth]{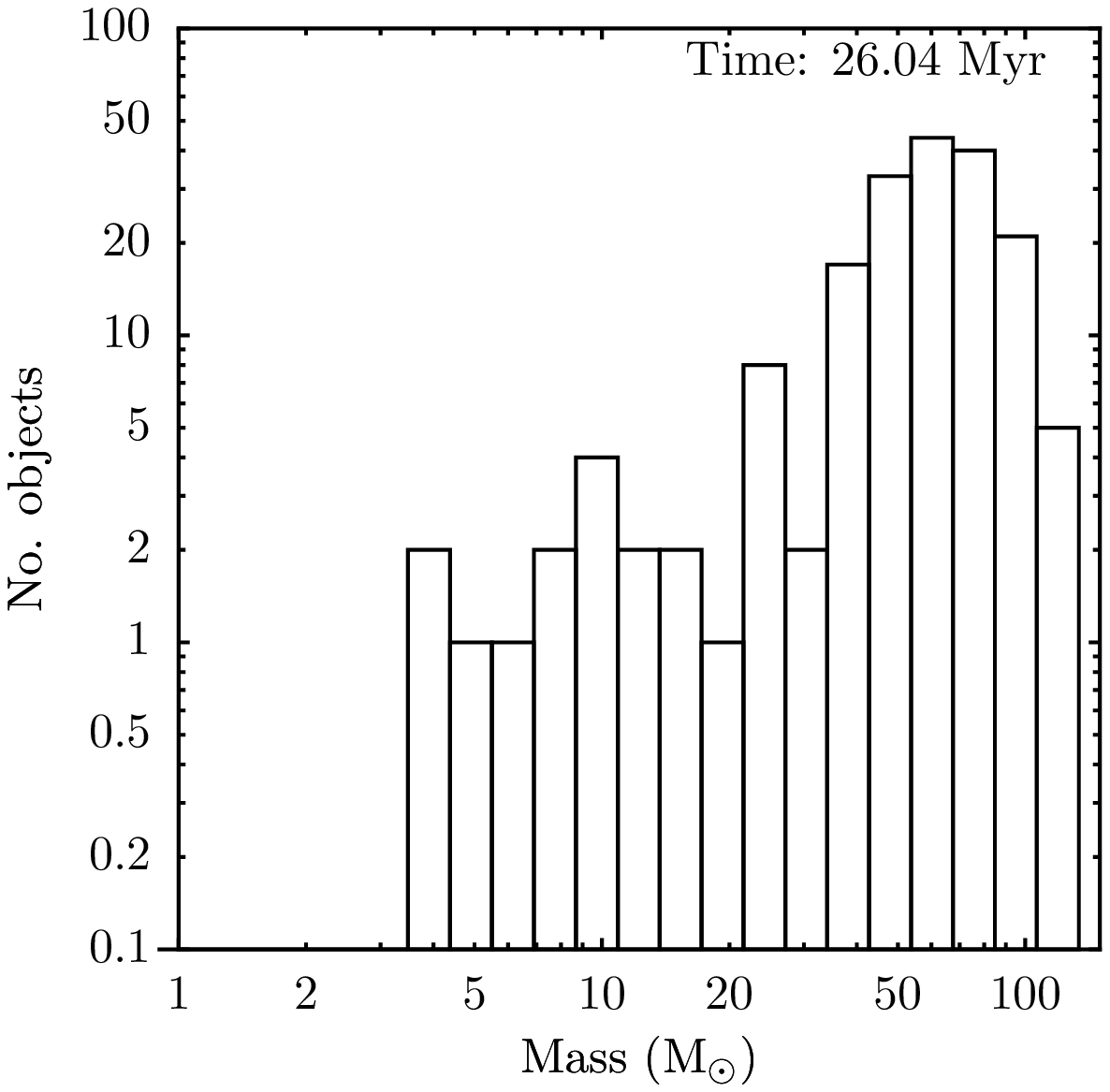}}
\vspace{0.1in}
\subfigure{\includegraphics[width=0.30\textwidth]{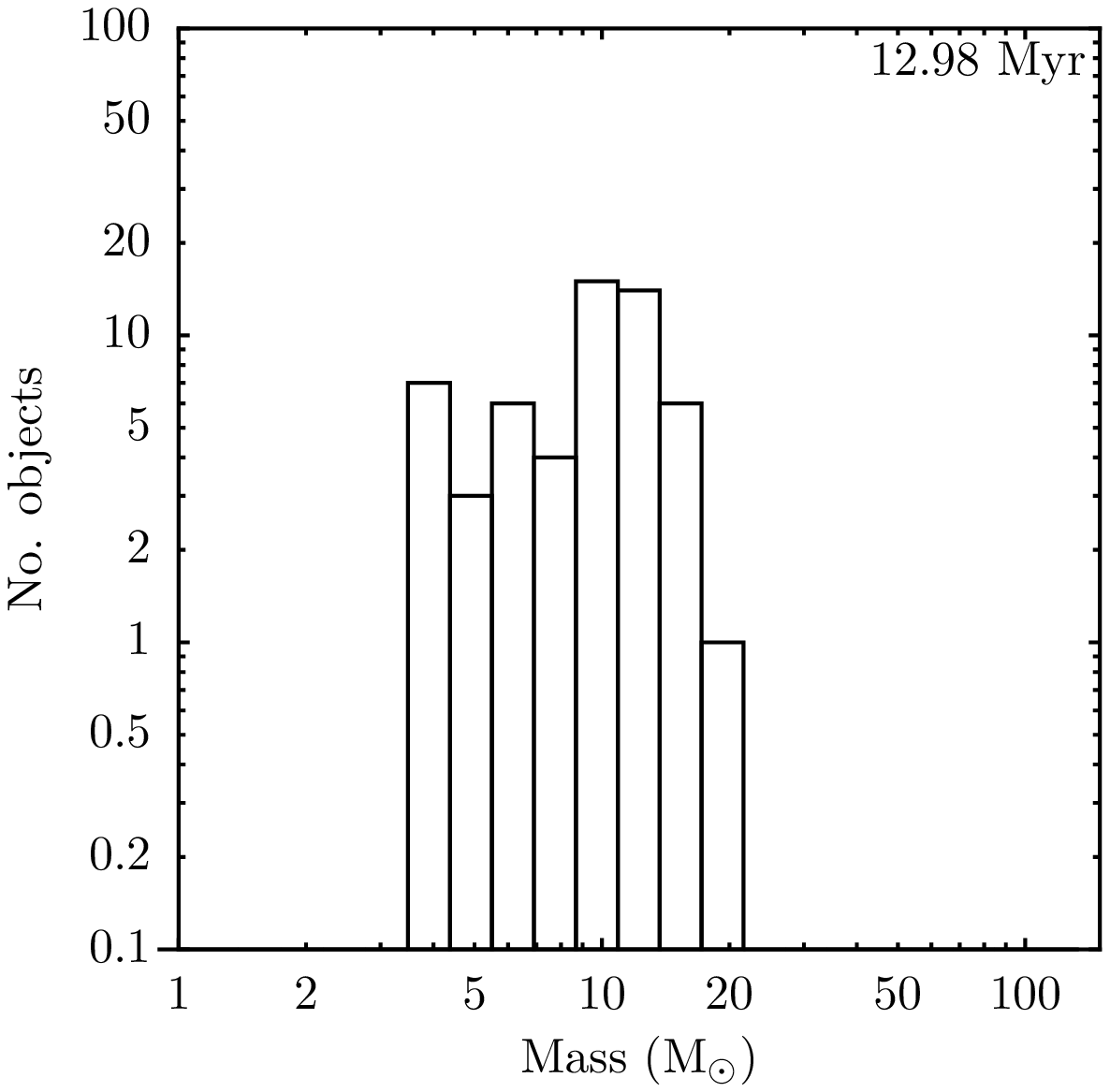}}
\hspace{0.1in}
\subfigure{\includegraphics[width=0.30\textwidth]{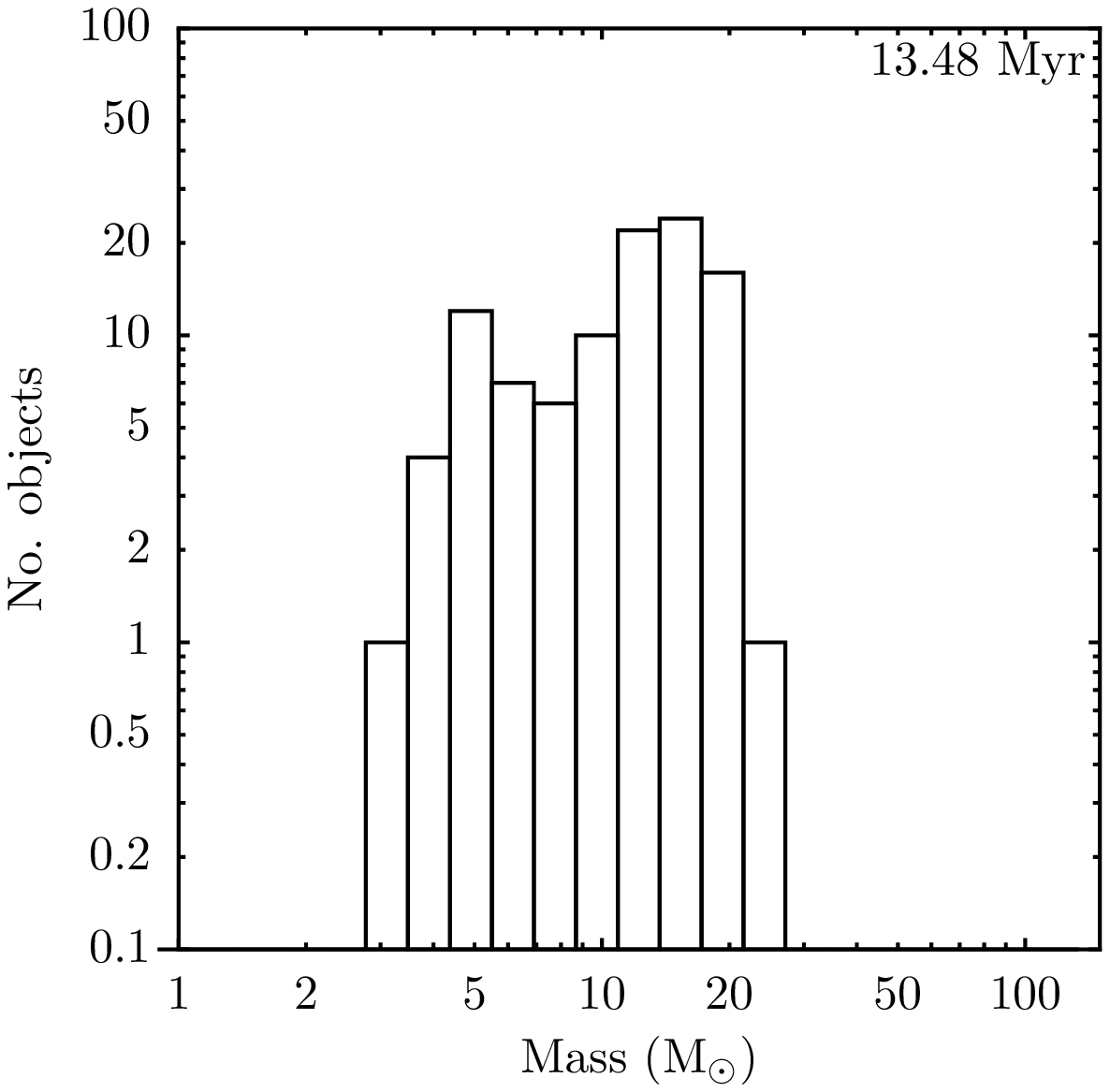}}
\hspace{0.1in}
\subfigure{\includegraphics[width=0.30\textwidth]{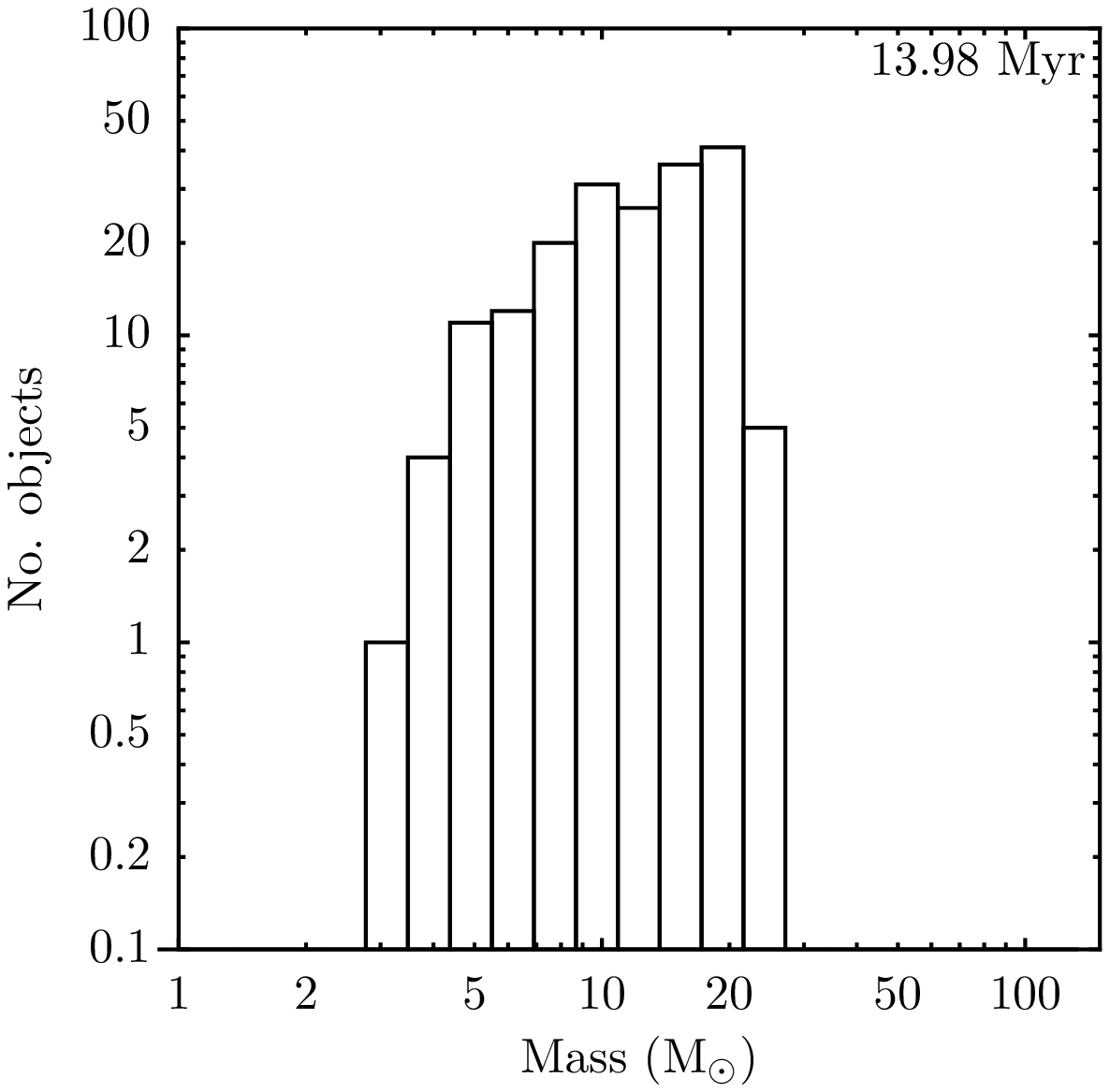}}
\caption{Mass functions of sink particles at three epochs in the medium--pressure calculation (top row) and the high--pressure calculation (bottom row).}
\label{fig:sink_mf}
\end{figure*}

\begin{figure}
\includegraphics[width=0.45\textwidth]{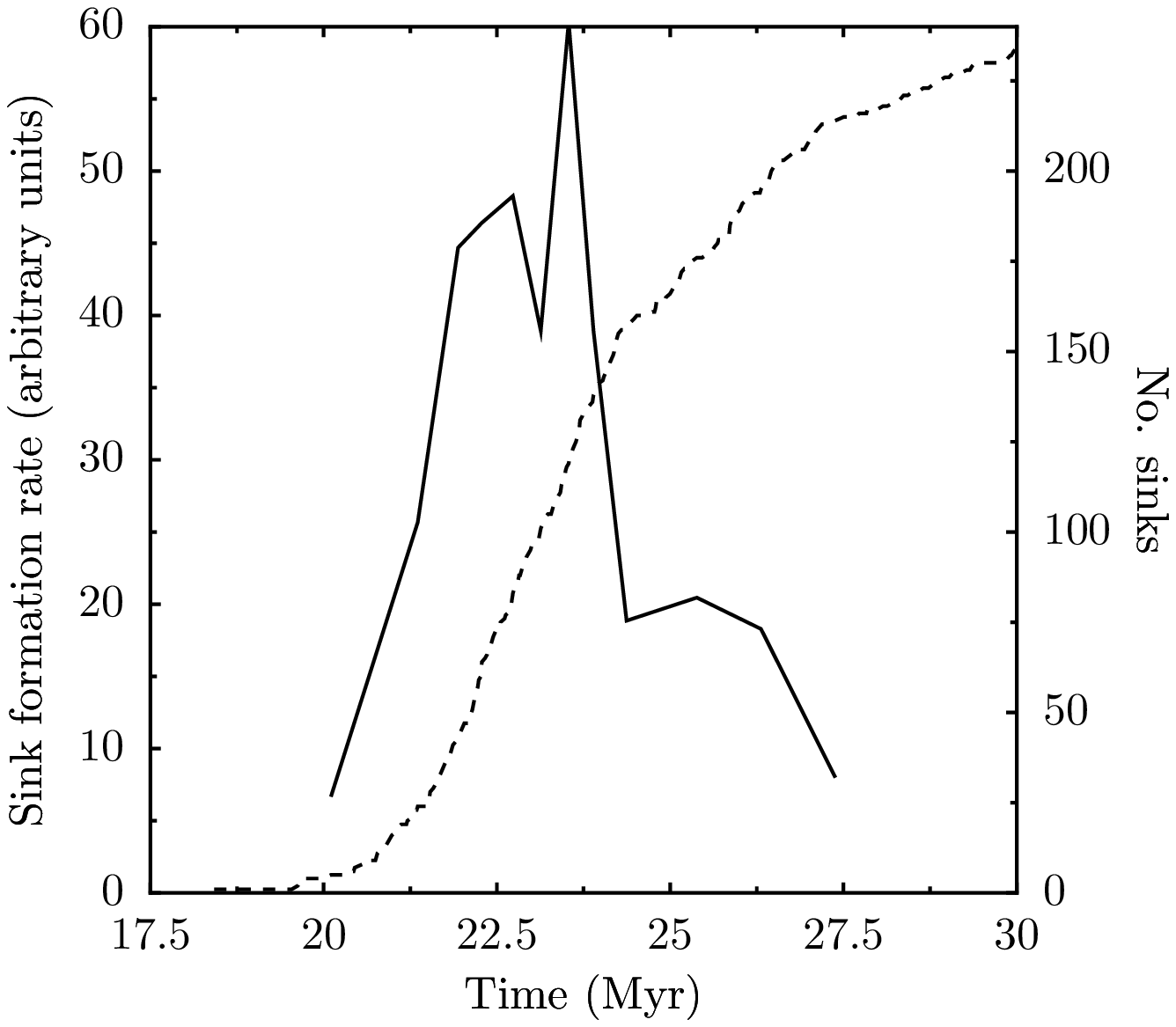}
\caption{Rate of formation (solid line) and total number (dashed line) of sink particles in the medium--pressure run plotted against time .}
\label{fig:sink_rate}
\end{figure}

\indent Once the p--cores start to become bound, they collapse and form sink particles and the mass function is more easily followed by tracing the masses of the sinks themselves, shown in Figure \ref{fig:sink_mf}. We see that the sink--particle mass functions rapidly come to resemble the clump mass functions presented in the previous section, since there are, by the epoch of sink--formation, very few clumps that are not bound and collapsing. The sink mass functions become ever more top--heavy, and even the high--mass end of the mass function loses its power--law slope. We note that the medium--pressure mass function peaks at a mass of $\sim80$M$_{\odot}$, corresponding to the mass associated with the most unstable wavenumber of $l\sim25$ from Figure \ref{hello}.\\
\begin{figure}
\includegraphics[width=0.45\textwidth]{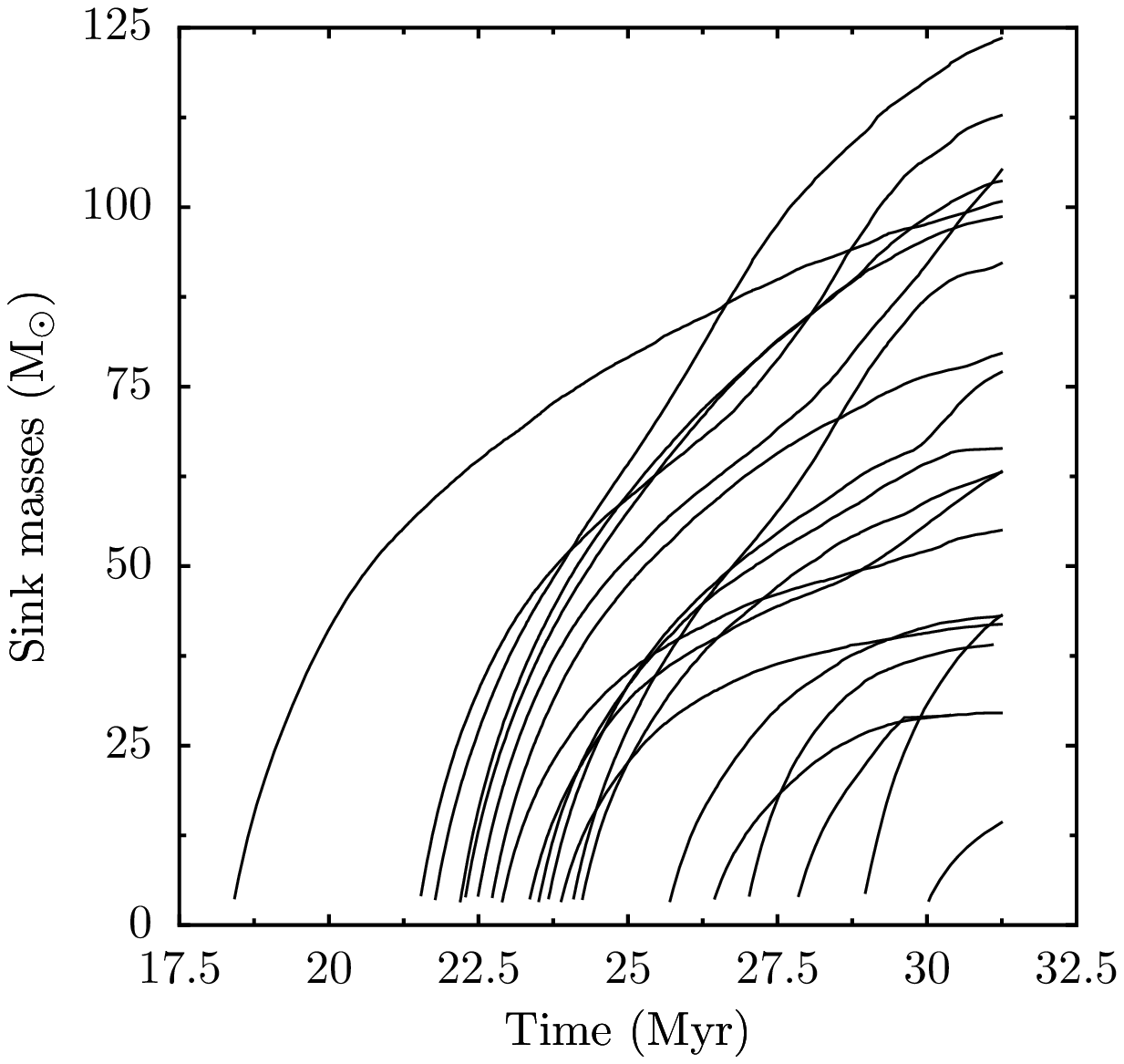}
\caption{Evolution of the mass of a randomly chosen 10$\%$ of the sinks formed in the simulation as a function of time in the medium--pressure calculation.}
\label{fig:sink_evolution}
\end{figure}
\begin{figure}
\includegraphics[width=0.45\textwidth]{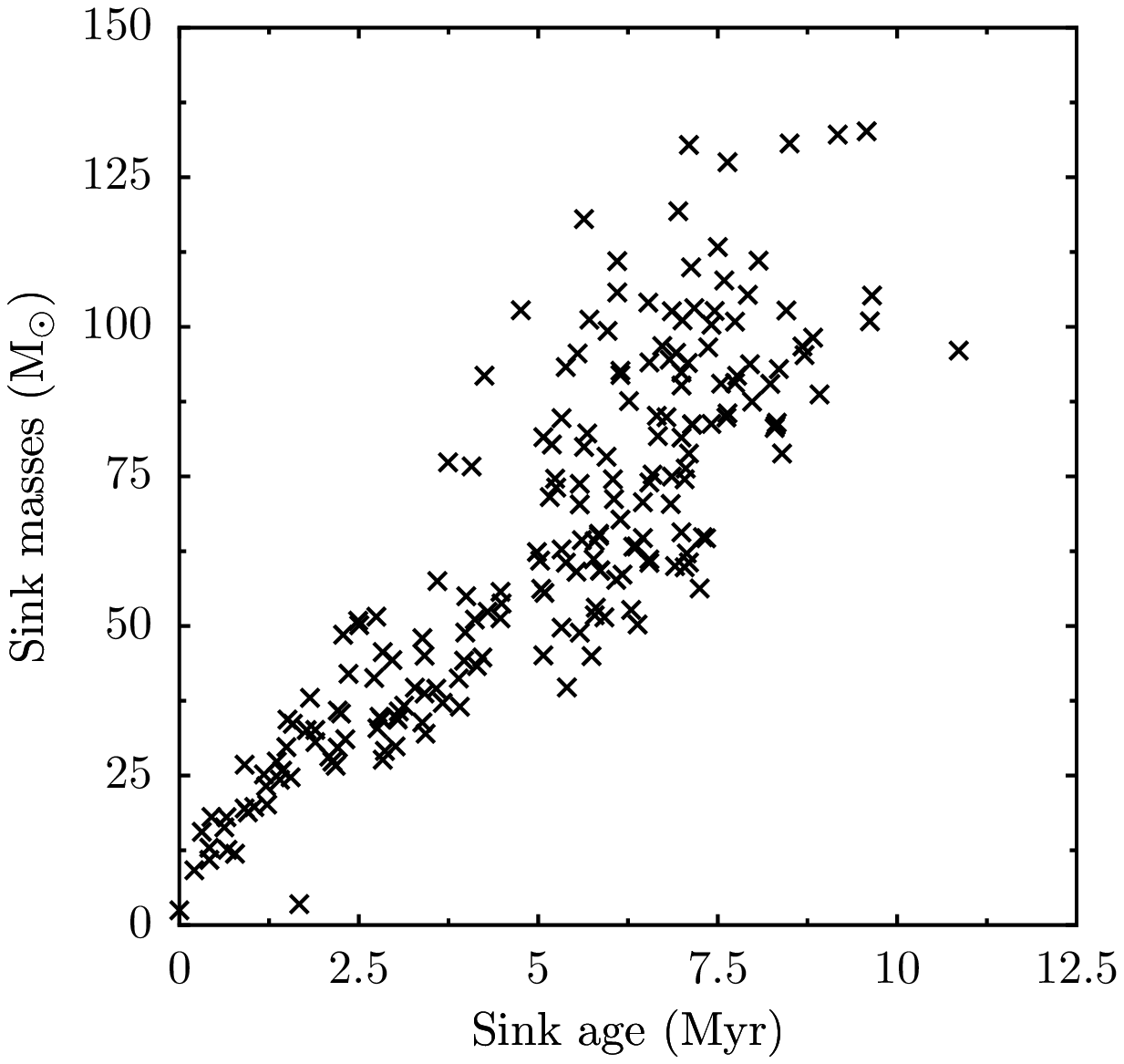}
\caption{Masses of sink particles as a function of their age as measured at a time of 31 Myr in the medium--pressure calculation.}
\label{fig:sink_mass_age}
\end{figure}
\indent Implicit in the thin--shell model is the assumption that a single perturbation will produce a single collapsed object whose mass is proportional to the mass of the perturbation, which in turn is given by $\lambda^{2}\Sigma/16$, where $\lambda$ is the perturbation wavelength and $\Sigma$ is the mean shell surface density at the time when the fragment collapses. In this picture, the accretion rate onto each collapsed object thus rises to a peak and falls back down to zero. On the contrary, as shown in Figure \ref{fig:sink_evolution}, we find that, although the accretion rate onto a given sink is a maximum at the time of the sink particle's formation and the accretion rate tails off, for most sink particles it never falls to zero and they continue accreting for the duration of the simulation so that, as shown in Figure \ref{fig:sink_mass_age}, sink masses are approximately proportional to their ages. We find that for most sink particles, the time--averaged  accretion rate is a strong function of neither age nor mass. The uniformity in the accretion rates of the sink particles is a consequence of the fact that most of them form in the period of time when the shell is almost stationary at its maximum radius, so that most sinks are born embedded in gas of the same density. Since the shell and the sinks are not in relative motion, the rate of accretion onto a given sink is determined only by the local freefall time, which is in turn the same throughout the shell at any given time.\\
\indent In Figure \ref{fig:sink_rate}, we show the rate of sink particle formation and the cumulative number of sink particles as functions of time in the simulation. The number initially rises strongly before beginning to tail off around 23 Myr. We showed in Figure \ref{fig:nclumps} that, as the first p--cores become bound, the total number of p--cores stabilises. The final number of sink particles in Figure \ref{fig:sink_rate} is very similar to the final number of p--cores, and the tailoff in the rate of sink--formation appears to be due simply to the fact that there are only a finite number of cores available from which to form sink particles. The top--heavy form of the mass function is then a consequence of the early pulse of sink--particle formation coupled with the fact that all sink particles share a common environment and thus accrete at comparable rates. We refer to this process as oligarchic accretion, since it is the sink particles that form first that accrete most of the shell's mass.\\
\section{Discussion}
The purpose of this paper was to simulate the fragmentation of an expanding shell, following the evolution well into the non--linear regime to see if the predictions of the thin--shell and PAGI models (nominally only valid in the linear regime) can be extrapolated to accurately determine the mass function of fragments formed. We find that the mass function of objects located by our clump--finding code agrees reasonably well with the predictions of the thin--shell model in the early stages of fragmentation before significant numbers of fragments become bound. We also find that the assumption implicit in the model that fragments do not interact appears to be sound. However, once p--cores start to become bound, the numbers of p--cores detected levels off at a number consistent with the most unstable wavenumber at this epoch. Once the first cores become bound, they suppress the formation of new fragments. Once sink particles begin to from from the p--cores, the rate of sink formation is initially high, but tails off as the available cores are consumed. Sink particles continue to accrete at roughly constant rates, since they all inhabit a similar environment, so that the initial pulse of sink--particle formation translates into a top--heavy mass function in which the mass corresponding to the most unstable wavelength at the time of the p--cores becoming bound is over--represented with respect to other modes. The analysis of \cite{1994MNRAS.268..291W}, in which the first unstable wavenumber is taken as representative of the mass function, describes fragmentation better than the analysis of  \cite{2001A&A...374..746W} in which all modes contribute to the mass function.\\
\indent We refer to this process as oligarchic accretion, since the objects that form first win simply by virtue of being first. This mechanism is very different from the competitive accretion described by \cite{2001MNRAS.323..785B}, since in that model it is the different environments of the few stars in the dense gas at a cluster centre that enable them to accrete more than their siblings. In the simulations presented here, most of the sink particles form in gas of the same density, around the time when the shell expansion stalls. It is instead the time of formation and the non--uniform rate of formation of sink particles which is important in determining the final mass of a given object. Those objects forming first are able to accrete more mass. The mass function becomes skewed because the rate of sink particle formation declines as the supply of p--cores formed during the linear phase of the shells evolution is consumed. \cite{2000ApJS..128..287K} observe a similar process in the evolution of a turbulent protocluster, where the mass function becomes skewed towards higher masses as the star formation efficiency becomes large and the gas reservoirs required to form new cores are depleted. The accretion process may also be aided by the fact that, at a time of $\approx$20 Myr, the shell begins to contract. However, the contraction is slow compared to the rate of sink particle formation and accretion, and the fragment mass function is already strongly top--heavy at this epoch, so contraction of the shell cannot be the primary driver of oligarchic accretion.\\
\indent It is possible that, in the case of a shell sweeping up mass as it expands, so that reserves of fresh gas in the shell are constantly replenished, p--cores would be able to form at all times, resulting in a mass function more closely resembling a power law. However, such a shell would experience the Vishniac instability \citep{1983ApJ...274..152V}, probably altering the mass spectrum of fragments in the non--linear portion of the shell's evolution. This is important, since our results suggest that the most unstable mode at the time when bound objects begin to form is over--represented in the mass function. To unequivocally demonstrate that accretion of fresh material would allow unabated p--core formation and produce a power--law mass function would require that the shell's gas reservoir be replenished in some way during sink formation. This is difficult in the case presented here, since most of the sinks form when the shell is almost stationary. The only way to replenish the gas would then be to actively feed matter into the shell, perhaps allowing it to infall from a reservoir just outside the shell's maximum radius, a highly artificial construction. We therefore defer answering this question to a later paper in which we analyse the effect of the Vishniac instability on a momentum--driven shell sweeping up an external medium.\\
\indent The mass functions produced by our simulations bear little resemblance to any known stellar or cluster mass function. As explained in Section 3, the mass resolution of our sink particles is good enough to resolve low--mass stars, but the linear resolution implied by the accretion radii (0.05pc) is rather large. Our sink particles should strictly be regarded as binaries or small multiple systems so that our mass functions, in the terminology of \cite{2008A&A...477..823G}, are Multiple Star Mass Functions (MSMFs), not Single Star Mass Functions (SSMFs). This would be true to an even greater degree if our results were rescaled to higher masses. \cite{2008A&A...477..823G} find, not surprisingly,  that the SSMF contains fewer high-mass objects and more low--mass objects relative to the MSMF, and that the SSMF peaks at lower masses. The SSMF from our simulations would probably therefore be slightly less deficient in low--mass objects and contain fewer high--mass objects than the mass functions we present, but the effects of multiplicity will not quantitatively change our finding that the mass functions become very top--heavy due to oligarchic accretion.\\
\indent \cite{2008PASJ...60.1297D} observed the mass function of CO clumps in the Carina Flare supershell and observed that their mass spectrum follows a power--law. Although they caution against over--interpretation of their data, their clump mass measurements imply that the clumps are not gravitationally bound (in the sense that their inferred masses are below their virial masses). This result is consistent with our findings that the as--yet--unbound objects produced in the linear phase of shell fragmentation have a power law mass function (although note that the power law we derive is considerably steeper than that inferred by \citet{2008PASJ...60.1297D}). Our findings suggest, however, that the mass function of these unbound fragments cannot necessarily be used to predict the mass function of stars or star clusters that will eventually form. There are as yet no observations of the stellar or cluster mass functions generated by the fragmentation of expanding shells, owing to the extreme difficulty of making such a measurement.\\
\indent The initial conditions used in our simulations, that of a shell which is pressure--confined and yet does not sweep up any mass are somewhat unusual and the peculiar mass function we obtain implies that most stars or clusters do not form in shells of this type. However, such shells must surely occur sometimes, since a shell may expand into the hot rarefied ISM and be pressure--confined while accreting negligible mass. Our results demonstrate that shells of this nature will produce anomalously high numbers of massive stars or clusters and may therefore lead to the propagation of star formation by triggering.
\section{Conclusions}
We find that, in the case of a momentum--driven shell pressure--confined internally and externally, the PAGI model faithfully reproduces the mass function of fragments produced by the gravitational instability during the time for which perturbations in the shell surface density are relatively small and the behaviour is linear.\\
\indent In the case studied here of a shell which does not sweep up mass, we find that a form of accretion which we term oligarchic accretion operates within the shell whereby the first objects to form become the most massive because they accrete from the gas reservoir for longer. Coupled with the finite number of p--cores formed during the linear fragmentation of the shell, this leads to a very top--heavy mass function in which the most unstable wavelength at the time when perturbations first become bound is over--represented. A shell whose gas reservoir is replenished may form fragments continuously, resulting in a mass function more closely resembling a power law.\\
\indent Since the mass function generated by our simulations is so unusual, we infer that shells of the type we have studied cannot make a strong contribution to the global stellar population, although no direct measurements of the mass functions produced by expanding shells exist against which we could test our conclusions.
\section{Acknowledgements}
The authors thank the referee, Sven Van Loo, for incisive and helpful comments which considerably improved the structure of the paper. JED acknowledges support from a Marie Curie fellowship as part of the European Commission FP6 Research Training Network `Constellation' under contract MRTN--CT--2006--035890. JED, RW and JP acknowledge support from the Institutional Research Plan AV0Z10030501 of the Academy of Sciences of the Czech Republic and project LC06014--Centre for Theoretical Astrophysics of the Ministry of Education, Youth and Sports of the Czech Republic. AW gratefully acknowledges the support of grant ST/H001530/1 from the
UK Science and Technology Facilities Council.
\section{Appendix A}
The thin shell dispersion relation
\begin{equation}
\label{omega}
\omega(l) = -\frac{AV}{R} + \sqrt{\frac{BV^2}{R^2} - \frac{c_s^2l^2}{R^2}
+ \frac{2\pi G \Sigma l}{R}}
\end{equation}
may be recast in a dimensionless form with the help of an expansion law -- a prescription for the variations in the shell radius $R$, expansion velocity $V$ and surface density $\Sigma$ with time.
The shells described in this work are ballistic and of fixed mass $M$, expanding due to their inertia into a vacuum and decelerated by self--gravity alone.\\
\indent We define a dimensionless radius $\xi$ as
\begin{equation}
\label{xi}
\xi = \frac{R}{R_\mathrm{max}} =  1 - \frac{K}{|W|} = 1 -
\frac{V^2R}{GM}
\end{equation}
where $K = MV^2/2$ and $W = (GM^2)/(2R)$ are the kinetic and potential energy respectively, and
\begin{equation}
\label{Rmax}
R_\mathrm{max} = \frac{R_0}{1-K_0/|W_0|}
\end{equation}
is the maximum radius to which the shell expands (the initial
total energy $K_0 + W_0$ at radius $R_0$ is assumed to be negative so that the shell is bound).
The free fall time obtained from Kepler's third law is
\begin{equation}
\label{tff}
t_\mathrm{ff} = \frac{\pi R_\mathrm{max}^{3/2}}{2(GM)^{1/2}}
\end{equation}
and the most unstable dimensionless wavenumber at $R_\mathrm{max}$ is
\begin{equation}
\label{lmax}
l_\mathrm{max} = \frac{\pi G \Sigma R_\mathrm{max}}{c_s^2} 
= \frac{GM}{4c_s^2 R_\mathrm{max}} \ .
\end{equation}
Inserting (\ref{xi}) and (\ref{lmax}) into (\ref{omega}) and normalizing it by
(\ref{tff}) we obtain
\begin{eqnarray}
w(l, \xi, l_\mathrm{max})&=& \omega(l, R, V, \Sigma, c_s) t_\mathrm{ff} \nonumber \\
&=&-\frac{A\pi}{2} (1-\xi)^{1/2} \xi^{-3/2} + \frac{\pi}{2}\times\nonumber \\
&& \left[ B(1-\xi)\xi^{-3} - \frac{l^2}{4 l_\mathrm{max}} \xi^{-2} +
\frac{1}{2}l\xi^{-3}\right]^{1/2}.
\end{eqnarray}
Using numerical factors $A = 3/2$ and $B = 1/4$ of a non-accreting shell we
obtain the final dispersion relation
\begin{eqnarray}
\label{dldisprel}
w(l, \xi, l_\mathrm{max}) &=& 
-\frac{3\pi}{4} (1-\xi)^{1/2} \xi^{-3/2} + \nonumber \\
&&\frac{\pi}{4}
\left[ \xi^{-3}(1+2l) - \xi^{-2}\left(1 + \frac{l^2}{l_\mathrm{max}}\right)
\right]^{1/2}.
\end{eqnarray}
The degree of fragmentation at a given wavenumber $l$ is described by the fragmentation integral
\begin{eqnarray}
\label{If1}
I_{f}(l,l_\mathrm{max}) &=& \int_{t_0}^{t} \omega(l,t',l_\mathrm{max}) dt' \nonumber \\
&=& \int_{\xi_0}^{\xi} \frac{w(l,\xi',l_\mathrm{max})}{t_\mathrm{ff}}
\frac{dt}{d\xi}d\xi'.
\end{eqnarray}
Since
\begin{equation}
\frac{d\xi}{dt} = \frac{1}{R_\mathrm{max}}\frac{dR}{dt} = \frac{V}{R_\mathrm{max}}
\end{equation}
and
\begin{eqnarray}
\frac{V}{R_{\rm max}} &=& (1-\xi)^{1/2}\left(\frac{GM}{R}\right)^{1/2}\frac{1}{R_{\rm max}} \nonumber \\
&=& (1-\xi)^{1/2}\left(\frac{GM}{R_\mathrm{max}}\right)^{1/2}\frac{1}{R_{\rm max}}
\left(\frac{R_\mathrm{max}}{R}\right)^{1/2} \nonumber \\
&=& \left(\frac{1-\xi}{\xi}\right)^{1/2} \frac{\pi}{2t_\mathrm{ff}},
\end{eqnarray}
where equations (\ref{xi}) and (\ref{tff}) have been used, we may write
\begin{equation}
\label{dtdxi}
\frac{dt}{d\xi} = \left( \frac{\xi}{1-\xi} \right)^{1/2} \frac{2t_\mathrm{ff}}{\pi}.
\end{equation}
Inserting (\ref{dtdxi}) into (\ref{If1}) yields the fragmentation integral in a form
\begin{eqnarray}
I_{f}(l,l_\mathrm{max}) & = & \frac{2}{\pi}\int_{\xi_{0}}^{\xi}
w(l,\xi',l_\mathrm{max})
\left( \frac{\xi'}{1-\xi'} \right)^{1/2} d\xi'
\nonumber\\
& = & -\frac{3}{2}\int_{\xi_{0}}^{\xi} \xi'^{-1} d\xi' + \frac{1}{2}\times\nonumber \\
&&
\int_{\xi_{0}}^{\xi}
\left[ \frac{1+2l}{\xi'^2(1-\xi')} - \frac{1+l^2/l_\mathrm{max}}{\xi'(1-\xi')}
\right]^{1/2} d\xi'.
\end{eqnarray}
Using the Sage symbolic manipulation package (available at \texttt{www.sagemath.org}), the fragmentation
integral becomes
\begin{eqnarray}
&&I_{f}(l,l_\mathrm{max}) = -\frac{3}{2}\ln\left(\frac{\xi}{\xi_0}\right)+\frac{1}{2}(2l+1)^{1/2}\times
 \nonumber \\
&&\ln\left\{
\frac{\xi}{\xi_0}\frac{2[(2l+1)l_\mathrm{max}]^{1/2} \alpha(\xi_0) + \beta\xi_0 + (4l+2)l_\mathrm{max}}
{2[(2l+1)l_\mathrm{max}]^{1/2} \alpha(\xi) + \beta\xi + (4l+2)l_\mathrm{max}}\right\}+\nonumber\\
&&\frac{1}{2}\left(\frac{l_\mathrm{max}+l^2}{l_\mathrm{max}}\right)^{1/2}\times \nonumber \\ 
&&\ln\left\{\frac{2(l_\mathrm{max}+l^2)^{1/2}\alpha(\xi_0) + (2l_\mathrm{max}+2l^2)\xi_0 + \beta}
{2(l_\mathrm{max}+l^2)^{1/2}\alpha(\xi) + (2l_\mathrm{max}+2l^2)\xi + \beta}
\right\},
\end{eqnarray}
where
\begin{equation}
\alpha(x) = [(l_\mathrm{max}+l^2)x^2+\beta x+(2l+1)l_\mathrm{max}]^{1/2}
\end{equation}
and
\begin{equation}
\beta = (-2l-2)l_\mathrm{max} - l^2.
\end{equation}

\bibliography{myrefs}

\label{lastpage}

\end{document}